\documentclass[manuscript]{aastex631}

\definecolor{Red}{rgb}{1,0,0}

\usepackage{bm}

\received{} 
\revised{} 
\accepted{} 
\submitjournal{ApJ}

\shorttitle{Modeling hot, anisotropic ion beams}
\shortauthors{Ofman et al.}

\begin{document}

\title{Modeling hot, anisotropic ion beams in the solar wind motivated by the Parker Solar Probe observations near perihelia}

\correspondingauthor{Leon Ofman}
\email{ofman@cua.edu}

\author[0000-0003-0602-6693]{Leon Ofman}
\affiliation{The Catholic University of America, Washington, DC 20064, USA}
\affiliation{Heliophysics Science Division, NASA Goddard Space Flight Center, 
Greenbelt, MD 20771, USA}
\affiliation{Visiting, Tel Aviv University, Tel Aviv, Israel}

\author[0000-0001-6018-9018]{Yogesh}
\affiliation{The Catholic University of America, Washington, DC 20064, USA}
\affiliation{Heliophysics Science Division, NASA Goddard Space Flight Center, 
Greenbelt, MD 20771, USA}

\author[0000-0002-5240-044X]{Scott A Boardsen}
\affiliation{Goddard Planetary Heliophysics Institute, 
 University of Maryland,  Baltimore, MD 21250, USA}
\affiliation{Heliophysics Science Division, NASA Goddard Space Flight Center, 
Greenbelt, MD 20771, USA}

\author[0000-0002-3808-3580]{Parisa Mostafavi}
\affiliation{Johns Hopkins University Applied Physics Laboratory, 
Laurel, MD 20723, USA}

\author[0000-0002-6849-5527]{Lan K Jian}
\affiliation{Heliophysics Science Division, 
NASA Goddard Space Flight Center, 
Greenbelt, MD 20771, USA}

\author[0000-0002-4001-1295]{Viacheslav M Sadykov}
\affiliation{Physics \& Astronomy Department, 
Georgia State University, 
Atlanta, GA 30303, USA}

\author[0000-0001-6038-1923]{Kristopher Klein}
\affiliation{Lunar and Planetary Laboratory, University of Arizona, Tucson, AZ 85721, USA}

\author[0000-0002-7365-0472]{Mihailo Martinovi\'{c}}
\affiliation{Lunar and Planetary Laboratory, University of Arizona, Tucson, AZ 85721, USA}






\begin{abstract}
Recent observations of the solar wind ions by the SPAN-I instruments on board the Parker Solar Probe (PSP) spacecraft at solar perihelia (Encounters) 4 and closer find ample evidence of complex anisotropic non-Maxwellian velocity distributions that consist of core, beam, and `hammerhead' (i.e., anisotropic beam) populations. The proton core populations are anisotropic, with $T_\perp/T_\parallel >1$, and the beams have super-Alfv\'{e}nic speed relative to the core (we provide an example from Encounter 17). The $\alpha$-particle population show similar features as the protons. These unstable VDFs are associated with enhanced, right-hand (RH) and left-hand (LH) polarized ion-scale kinetic wave activity, detected by the FIELDS instrument. Motivated by PSP observations, we employ nonlinear hybrid models to investigate the evolution of the anisotropic hot-beam VDFs and model the growth and the nonlinear stage of ion kinetic instabilities in several linearly unstable cases. The models are initialized with ion VDFs motivated by the observational parameters. We find rapidly growing (in terms of proton gyroperiods) combined ion-cyclotron (IC) and magnetosonic (MS) instabilities, which produce LH and RH ion-scale wave spectra, respectively. The modeled ion VDFs in the nonlinear stage of the evolution are qualitatively in agreement with PSP observations of the anisotropic core and `hammerhead' velocity distributions, quantifying the effect of the ion kinetic instabilities on wind plasma heating close to the Sun. We conclude that the wave-particle interactions play an important role in the energy transfer between the magnetic energy (waves) and random particle motion leading to anisotropic solar wind plasma heating.
\end{abstract}



\section{Introduction} \label{intro:sec}
The extended heating of the solar wind plasma in the inner heliosphere is still not well understood. It is widely recognized that the dissipation of magnetized fluctuations within the solar wind occurs at small kinetic scales, specifically below the 'break point' of the large-scale, low-frequency magnetohydrodynamic (MHD) turbulent cascade, as noted in recent studies \citep[see, e.g.,][]{Che14,Vec18,Bow20,Bow24,Ale21}. This dissipation takes place at frequencies that are near the proton gyro-resonance, i.e., ion-kinetic scales, where $\alpha$ particles can play a significant role \citep[see, e.g., recently,][]{Ofm23,McM24,Xio24,Mar24}. The energy of large-scale fluctuations is transferred to these smaller scales through larger-scale phenomena, such as turbulence cascades, wave interactions, and magnetic reconnection. Recently, it has been demonstrated through hybrid-PIC modeling how the large scale turbulent Alfv\'{e}nic fluctuation could cross the so-called  `helicity barrier' and supply energy to generate ion-scale waves such as ion-cyclotron waves (ICWs) that lead to anisotropic ion heating \cite{Squ22,Squ23}. On kinetic scales, instabilities and resonances, lead to conversion of magnetic and kinetic energy into the thermal energy of the nearly collisionless plasma through wave-particles interactions. Since the solar wind plasma is nearly collisionless, the steady state is not necessarily Maxwellian but is limited by the stability thresholds of various kinetic instabilities (such as temperature anisotropy driven instabilities commonly visualized as a `Brazil' plot of solar wind data evident in the temperature anisotropy and $\beta_\parallel$ parameters space, e.g., \citet{Bal09,Maru12,Che16,Yoo17,Mart23,Yoo24}). Although collisions can influence the solar wind plasma over extensive distances and prolonged timescales, characterized by quantities such as the collisional age \citep[e.g.,][]{Neu76,Kas08, Mos24_col}, the impact of collisions is negligible in the acceleration region of the fast solar wind close to the Sun (beyond several solar radii), characterized as `young solar wind'. Evidence from historical Helios data, as well as recent observations from the Parker Solar Probe (PSP) \citep{Fox16} and Solar Orbiter (SolO) \citep{Mul20}, indicates that proton velocity distribution functions (VDFs) are significantly non-Maxwellian and display characteristics of super-Alfv\'{e}nic beams accompanied by kinetic wave activity \citep[e.g.,][]{Mar82a,Mar82b,Mar12,Ver20,Lou21}. The wealth of in situ data from the SWEAP/SPAN-I \citep{Kas16,Liv21} and FIELDS \citep{Bal16} instruments on board the PSP spacecraft provides valuable information about proton and $\alpha$ particle populations in the solar wind with ample evidence of non-Maxwellian velocity distributions and associated ion-scale kinetic wave activity, in particular during PSP perihelia that lead to many discoveries \citep[see e.g., the review][]{Rao23}. The unprecedented new detailed data of solar wind ion VDFs and ion-scale kinetic wave activity very close to the Sun demonstrate that ion kinetic instabilities are important components in the heating and energization of the solar wind plasma.

Recently, \citet{Ver22} found in data from PSP's fourth perihelion (Encounter 4, at about 28$R_s$) anisotropic super-Alfv\'{e}nic proton `hammerhead'-shaped beams with speeds of $\gtrsim 2.5$ times the local Alfv\'{e}n speed, while super-Alfv\'{e}nic $\alpha$ particle beams were also detected \citep{McM24}. During more recent perihelia passages (Encounters 8-9, at about 16$R_s$), PSP has identified extended periods of sub-Alfv\'{e}nic solar wind, where the solar wind speed is lower than the local Alfv\'{e}n speed, facilitating incoming wave interactions, first reported by \citet{Kas21} with further analysis by \citet{Ban22}. These conditions are particularly relevant for advancing the understanding of solar wind plasma instabilities, acceleration mechanisms, and local heating processes. 

The nonthermal features in the ion VDFs such as beams and anisotropies can drive kinetic instabilities and provide a significant source of free energy that can result in solar wind plasma heating and acceleration \citep[e.g.,][]{Gar00,Mar06,Bou13,Ver13,Ver19,OVM14,Kle18,Mar21,Ofm22a,Ofm23,Ver22,Wal23,Shaa24,McM24}. The PSP observations show that ion beams and ion-scale kinetic waves play an important role in the dynamics and energetics of the young solar wind. Using these new data, kinetic instabilities, as well as the acceleration and heating of the solar wind, have recently been studied in detail \citep[e.g.,][]{Bow20,Bow24a,Bow24b,Ver20,Vec21,Kle21,Sha24,McM24}. Recently, \citet{Pen24} studied the preferential heating and acceleration of $\alpha$-particles in the solar
wind, observed by PSP and found  correlation between  $\alpha$-particles and proton drift velocity and the corresponding temperature ratios $(T_\alpha/T_p$) in the young solar wind close to the corona. The instabilities and heating due to $\alpha$-proton drift velocity were investigated extensively in our previous hybrid modeling studies \citep[e.g.,][]{XOV04,MVO13,MOV15,OVM14,OVR17} with similar conclusions on the positive correlation between super-Alfv\'{e}nic $\alpha$-proton drift velocity and $\alpha$ particle heating.

The hybrid models, in which ions are modeled using the Particle-In-Cell or PIC method, and electrons are modeled as fluids \citep{WO93}, have been extensively developed in the past to study the heating and ion kinetic instabilities of multi-ion SW plasma in 2.5D and 3D \citep[e.g.,][]{Gar01,Gar03,Gar06,HT06,OV07,Ofm10a,OVM11,OVM14,OVR17,OOV15,MOV15,Ofm19b,Mar20,Ofm22a,MV24}.  The ion instability-produced wave spectra can heat the ions by resonant interactions, as well as through non-resonant parametric decay instabilities of Alfv\'{e}n waves in the collisionless kinetic regime \citep[e.g.,][]{Ara07,VM11,Ver12} and stochastic heating \citep[e.g.,][]{BC13,Vec17}. 

Recently, \citet{Ofm22a, Ofm23} motivated by PSP/SPAN-I and FIELDS instrument observations near perihelia, modeled ion beams, kinetic instabilities, and associated waves using 2.5D and 3D hybrid models. Their modeling results demonstrated the growth, nonlinear saturation and damping of ion kinetic instabilities in the solar wind plasma, the temporal evolution of the ion beams and temperature anisotropies. The studies demonstrated the partition of energies between the ions and fields in the non-linear regime, and concluded that the ion-beam-driven kinetic instabilities can play an important role in the dissipation of energy on kinetic scales, impacting solar wind heating. Fully kinetic (PIC) 2.5D modeling study of proton beam driven instability for other beam parameters motivated by PSP observations found qualitatively similar result \citep{Pez24}. The previous  study by \citet{Ofm22a} was focused on cool (low-$\beta$) initial state, with initially isotropic ion temperatures{, motivated by the observations at Encounter 4.}. The PSP data at recent perihelia demonstrated that the VDFs cores and beams could become anisotropic and the `hammerhead'-shaped beam VDS are consistent with hot anisotropic proton and $\alpha$ beams \citep{Ver22,Ofm23,McM24,Shi24}. Thus, new, closer PSP encounter data became available, providing motivation for new Hybrid-PIC studies of the solar wind plasma, and here we use PSP  data from Encounter 17 to motivate the modeling for our present study, that investigates hot anisotropic ion beams, associated kinetic wave activity, instabilities and anisotropic heating.

The paper is organized as follows, Section~\ref{obs:sec} is devoted to the observational motivations for the present study, based primarily on PSP/SPAN-I and FIELDS instrument data.  Section~\ref{model:sec} describes the hybrid model employed in the present study, with the numerical results in Section~\ref{num:sec}. Finally, the discussion and conclusions are given in Section~\ref{disc:sec}

\section{Observational Motivations} \label{obs:sec}

Observations show that VDFs with strong beams and hammerheads are often linked to intense ion-scale waves. Recent studies by \cite{Ver20} demonstrated the association between proton beams and observed ICWs, while \cite{Ver22} showed that hammerheads are typically associated with kinetic-scale waves. Hammerheads exhibit highly anisotropic VDFs, which can serve as a sources of free energy. The release of this energy, as the ion distribution becomes more isotropic through wave-particle interactions, can generate ion-scale waves depending on the degree of anisotropy, $\beta_\parallel$, and other stability parameters in the distribution. More recently, Yogesh et al. 2025, (in prep) showed that the occurrence and frequency band of these waves change with the density of observed ion beams.
\begin{figure}[ht]
\centering
\includegraphics[width=0.65\linewidth]{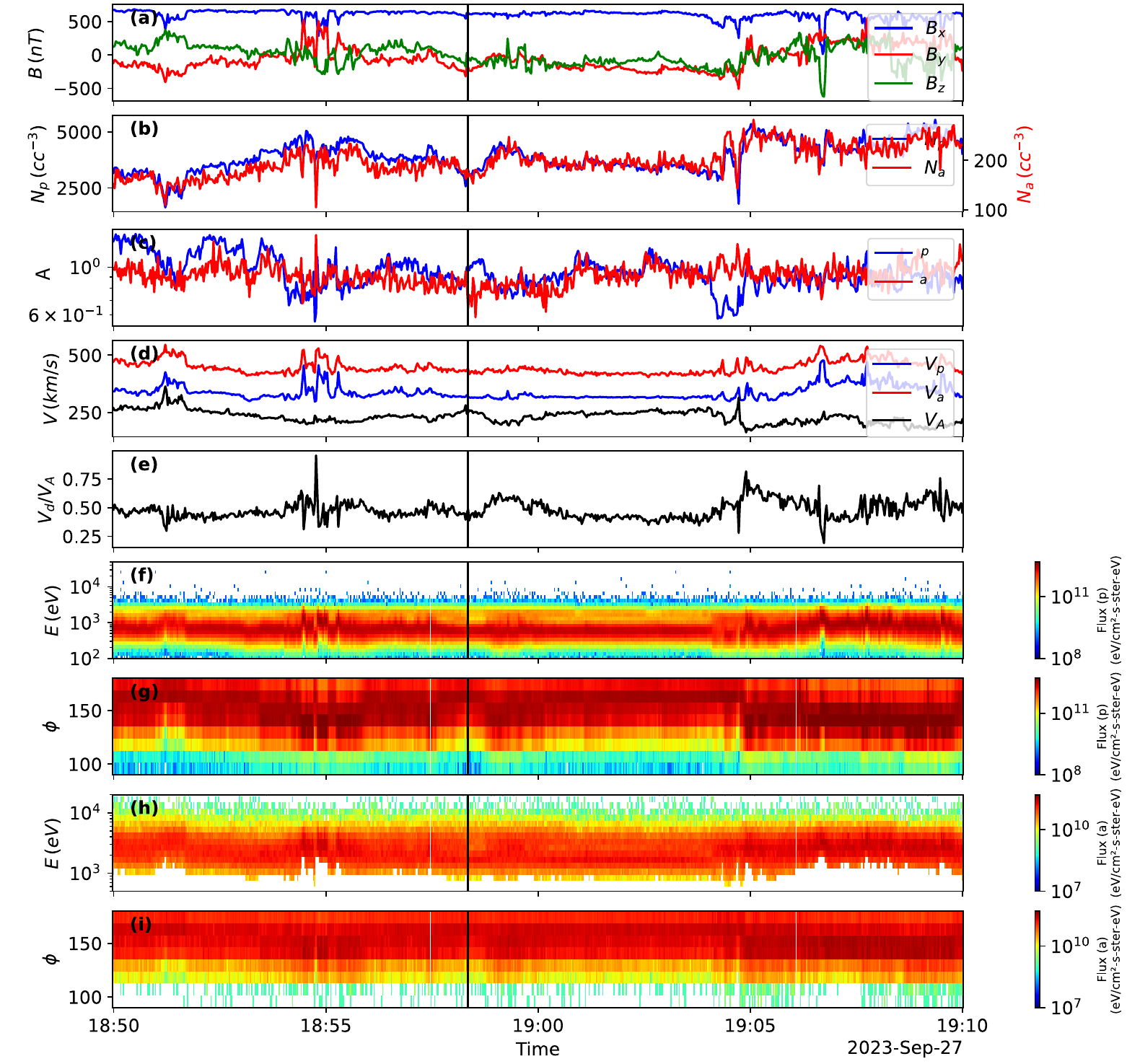}\\
\includegraphics[width=0.6\linewidth]{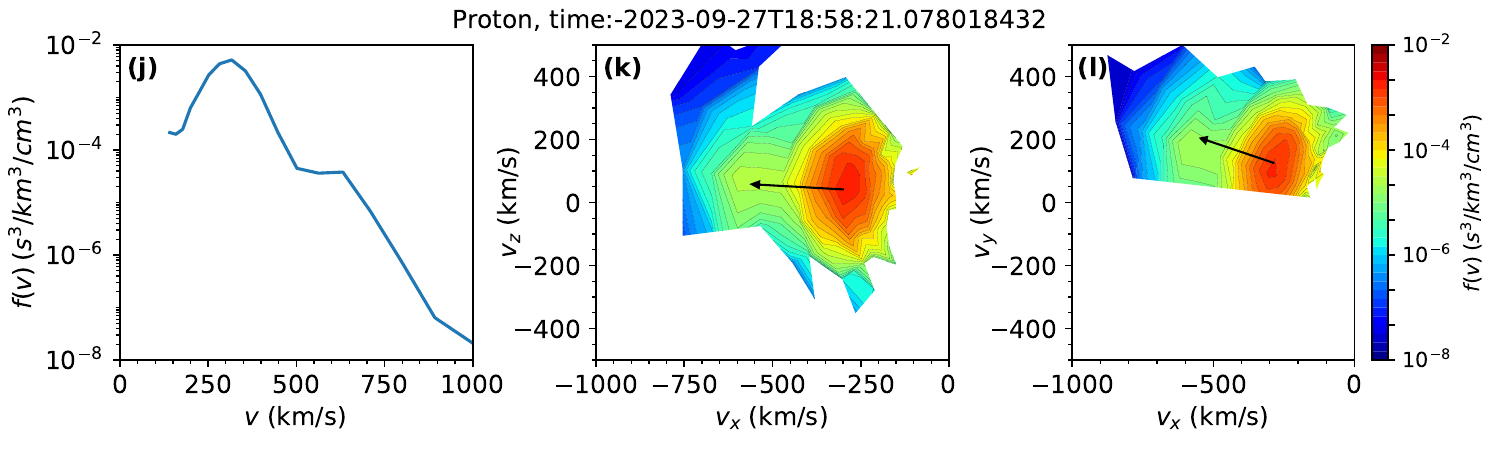}\\
\includegraphics[width=0.6\linewidth]{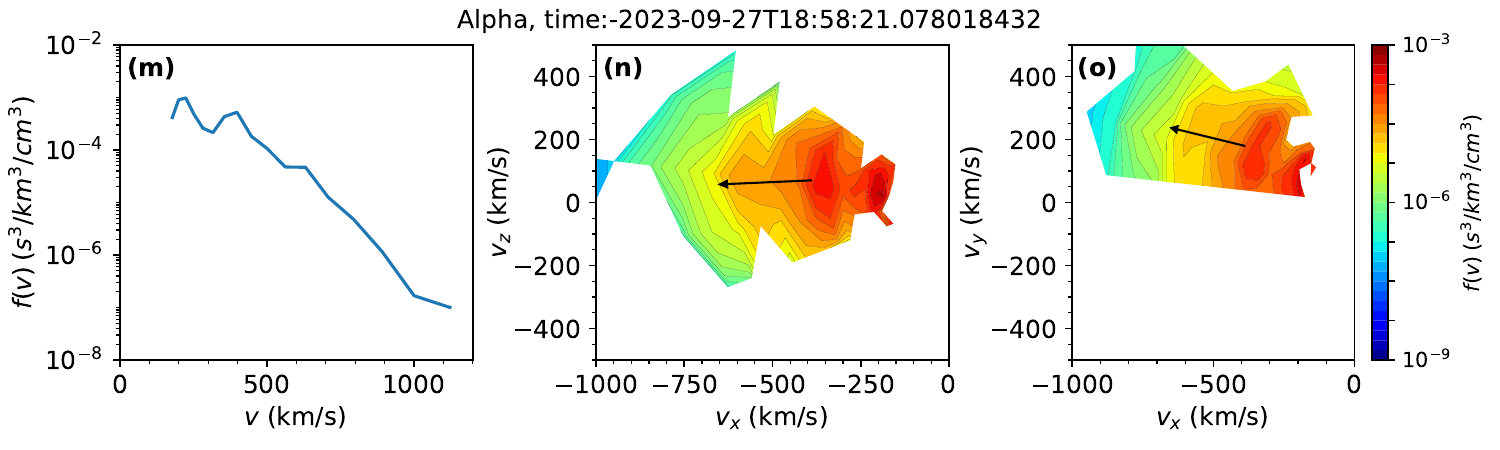}
\vspace{-0.5cm}\caption{\small The temporal evolution at Encounter 17 of (a) magnetic field components, (b) proton number density (blue) and $\alpha$ particle number density (red), (c) temperature anisotropy ($A = T_\perp/T_\parallel$) of protons (blue) and $\alpha$ particles (red), (d) velocity of protons (blue), $\alpha$ particles (red), and Alfv\'{e}n velocity (black), (e) ratio of the differential velocity between $\alpha$ particles and protons to the Alfv\'{e}n velocity, (f) proton flux variation with energy, (g) proton flux variation with azimuthal angle, (h) $\alpha$ flux variation with energy, and (i) $\alpha$ flux variation with azimuthal angle are shown in panels. The vertical black line shows the time of the VDFs of the protons and $\alpha$ particles in lower panels. Panels (j)$\&$(m) show the VDF summed over all look directions, (k)$\&$(n) show the VDF contours summed and collapsed onto the azimuthal plane, (l)$\&$(o) show the VDF contours summed and collapsed onto the $\theta$ plane. The black arrow represents the magnetic field direction in SPAN-I coordinates, where the tail is at the peak of VDFs and the length is the Alfv\'{e}n speed in km/s.}
\label{vdfs:fig}
\end{figure}

In this section, we show some observational examples from PSP data that provide motivation for the hybrid modeling study. We utilize measurements from the SPAN-I instrument\citep{Liv22}, part of the Solar Wind Electron Alpha and Protons (SWEAP) investigation \citep{Kas16}. The SPAN-I instrument comprises a time-of-flight section and an electrostatic analyzer, designed to capture the 3D VDFs of ions within an energy range of 2 eV to 30 keV in the solar wind (SW). SPAN-I can measure the bulk solar wind only when the VDF peak enters its field of view (FOV), which occurs during limited periods around perihelia \citep{Mos_22}. The partial observed VDF because of FOV issue can be seen in Figure 1.A of \cite{Woodham2021}. We carefully selected appropriate time intervals to analyze the VDFs observed by PSP that minimize the adverse effects of the limited FOV.
In Figure~\ref{vdfs:fig}, we show an example of the time interval that includes two populations (core and beam) for protons and $\alpha$ particles, observed by PSP on 2023 September 27 UT during Encounter 17 at a heliocentric distance of 0.06 au or 13 solar radii ($R_s$). 


Figure~\ref{vdfs:fig} shows an example of super-Alfv\'{e}nic beams of protons and $\alpha$ particles from PSP encounter 17. The Alfv\'{e}nic velocity ($V_A=B/\sqrt{\mu_0(n_p m_p+n_\alpha m_\alpha)}$) corresponding to the interval shown in Figure~\ref{vdfs:fig} is $\sim$ 249 km/s. The initial calculation from the 1D VDF plot shows that the proton beams have a drift velocity of $\sim$ 1.3$V_A$ with respect to the proton core. The $\alpha$ particle VDFs show a three-part structure including proton leakage at low energy (i.e., instrumental effect), $\alpha$ core, and $\alpha$ beams. The proton leakage in $\alpha$ particle VDFs is discussed in Section 6.2 of \citet{Liv22} and in \citet{McM22}. The initial calculation from 1D plot shows that the $\alpha$ particle beams have a drift velocity of $\sim$ 1.03$V_A$ with respect to the proton core. The lower energy peak in the $\alpha$s is not real but due to proton contamination of the $\alpha$ particle  measurements. However, due to observational limitations, the $\alpha$ particle beam speed value is difficult to determine from the observed VDFs.
\begin{figure}[ht]
\centering
\includegraphics[width=0.8\linewidth]{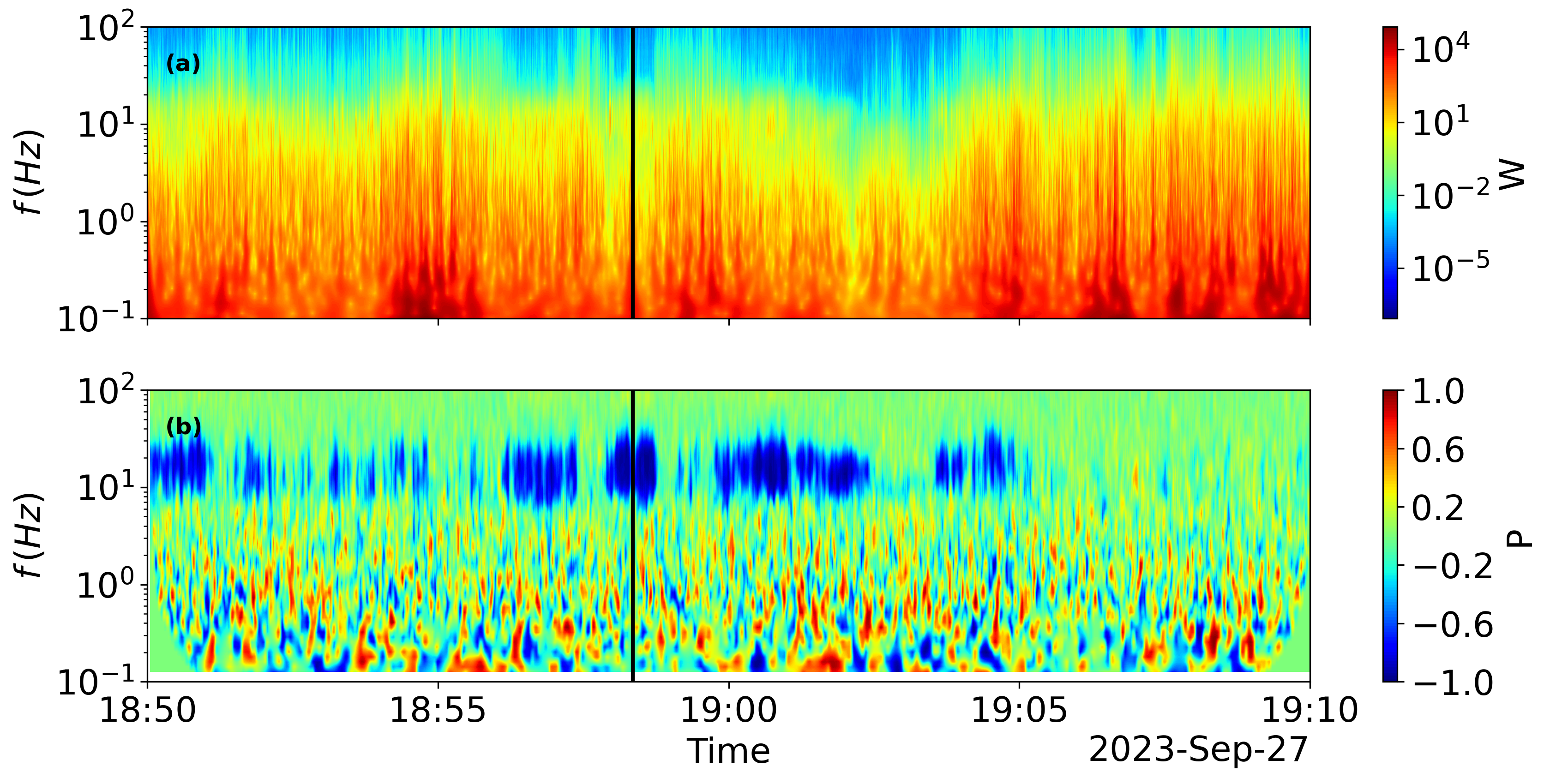}
(c)\includegraphics[width=0.55\linewidth]{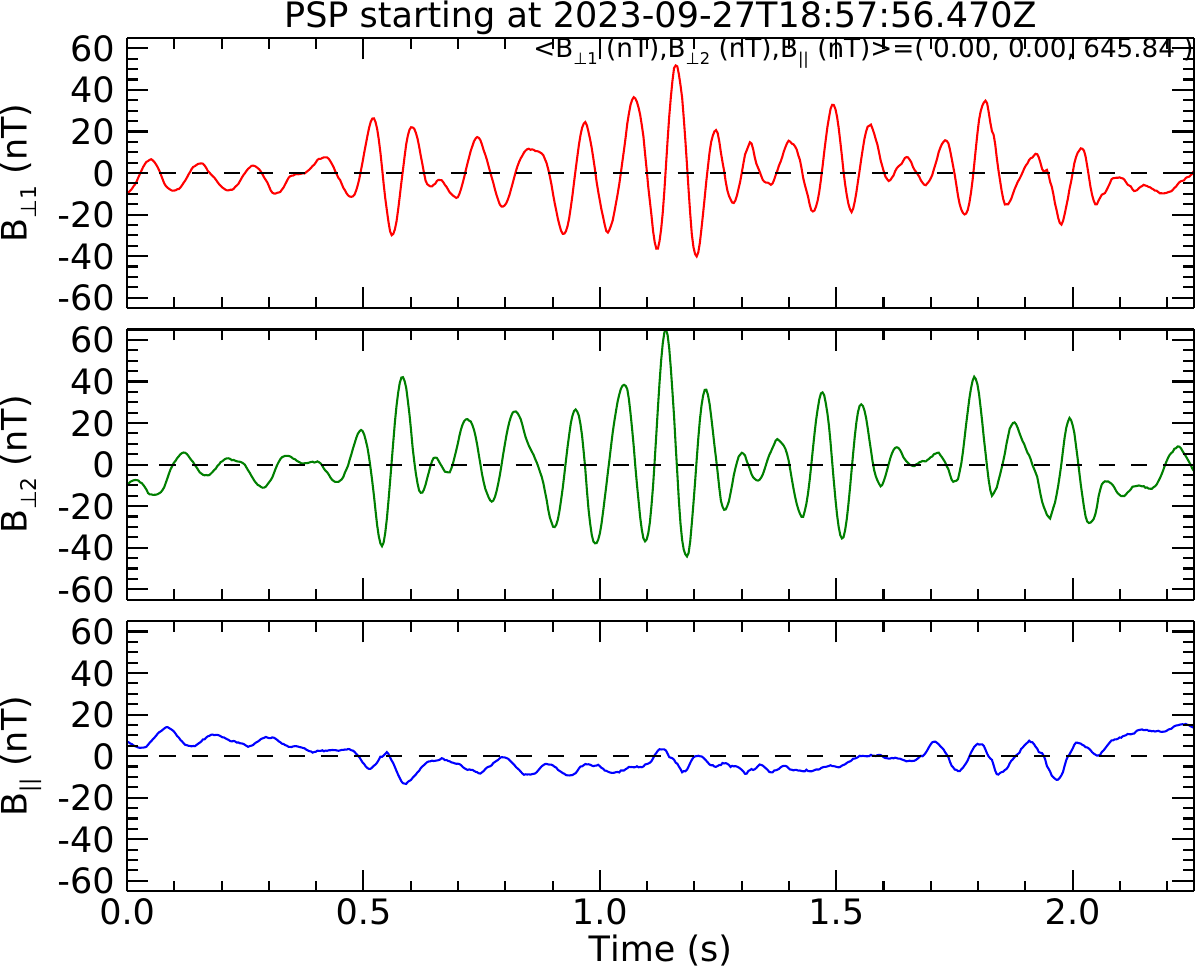}
(d)\includegraphics[width=0.35\linewidth]{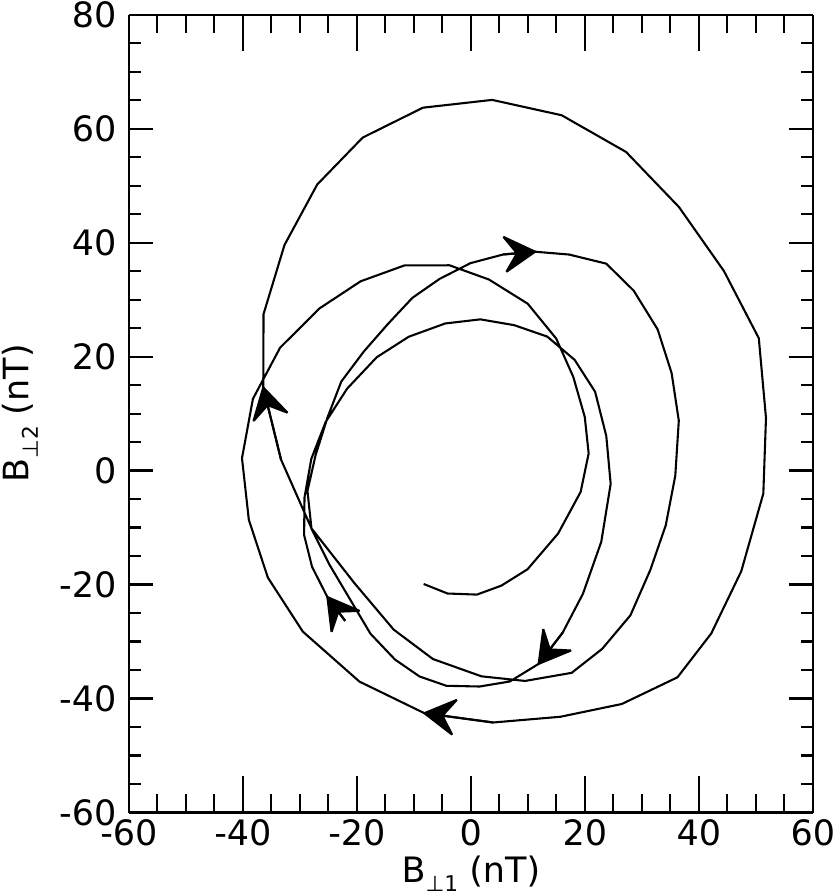}
\caption{ The wavelet power spectra (W, panel-a) and polarization (P, panel-b) for the same time interval shown in Figure~\ref{vdfs:fig}. The left-handed polarized ion-scale waves associated with the observation of the beams are evident in the dark-blue patches in panel (b). (c) Temporal sub-interval showing ICWs in field aligned coordinates, ($B_{\perp 1}$, $B_{\perp 2}$, $B_\parallel$) form a right handed coordinate system, the 646 nT background field has been subtracted, (d) magnetic field hodogram in the perpendicular plane of the four dominant wave cycles in (c) where the arrows indicate the direction of rotation which is left-handed.}
\label{waves}
\end{figure}

Figure~\ref{waves} shows an example of ion-scale wave activity analysis based on the measurements of the magnetic field by the fluxgate magnetometer of the PSP FIELDS suite during the time interval 2023 September 18:50 UT to 19:10 UT, the same interval shown in Figure \ref{vdfs:fig}. The details for computing the wave power spectrum and the polarization of the observed waves can be found in \citet{Shankarappa2023}. The black vertical line in Figure~\ref{waves} represents the same time interval for which $\alpha$ particle and proton VDFs are shown in Figure~\ref{vdfs:fig}. Left-handed ion-scale wave observations can also be seen as dark blue patches during this interval. Polarization analysis \citep[i.e.,][]{Boa16} was performed over this interval; filtering this analysis by 6 Hz $<$ frequency $<$ 20 Hz and the degree of polarization $>$ 0.8 and the ellipticity $< -0.5$ and the power spectral density $> 1$ $nT^2/Hz$; the wave normal angle is estimated to be $4.3^\circ\pm 2.9^\circ$ folded into the first quadrant and the ellipticity is $-0.87\pm0.08$ for these left-handed waves, an ellipticity of 1, 0, -1 is right-handed circularly, linear, left-handed circularly polarized, respectively. 

Time series of a sub-interval is shown in Figure~\ref{waves}c in field aligned coordinates. Strong perpendicular coherent left-handed circularly polarized oscillations are observed with a period near the proton cyclotron frequency which is near 10 Hz. The magnetic field hodogram in Figure~\ref{waves}d shows the left-handed polarization of the four dominant wave cycles in Figure \ref{waves}c. The perpendicular (transverse) wave power dominates the parallel (compressional) power. 

These waves are consistent with Doppler shifted (spacecraft frame relative to the solar wind flow frame) ICWs. To estimate the Doppler shift the average plasma parameters are used for the time interval of Figure~\ref{waves}a: $N_p = 3300$ cm$^{-3}$ and $N_\alpha=180$ cm$^{-3}$, background magnetic field magnitude of 646 nT, wave normal angle of $0^\circ$, and a relative flow between the bulk proton velocity and spacecraft frame along background magnetic field of 315 km s$^{-1}$. Using these parameters, we solve for the Doppler shift from the spacecraft to the solar wind flow frame using the left-handed branch in cold plasma dispersion relation (ignoring hot plasma effects). Using 10 Hz for the spacecraft frame frequency, we get two  possible solutions for the solar wind flow frame frequency of 3.6 Hz ($\alpha$ ICW branch) and 5.3 Hz (proton ICW branch) with wavelengths of 49 and 68 km, respectively. 

The hybrid simulation results presented in the following sections, which are qualitatively based but not intended to model the specific observations in detail. Units for the hybrid simulation are dimensionless, to help translate between observation and simulation, using the average parameters, wavenumber ($k$) in the unit of inverse proton inertial length $\delta_p^{-1}=\omega_{pp}/c$ or 0.26 km$^{-1}$, angular frequency in the unit of $\Omega_p$ or 62 radians/s based on the above observational values. 


The association of strong beams and ion-scale waves in the observations suggests that the waves are produced by the ion kinetic instabilities, and that wave-particle interactions are occurring. Using these observations as motivational input, we aim at reproducing the waves generated by the kinetic instabilities in the proton and $\alpha$ particle VDFs with the hybrid model (see Sections~\ref{model:sec}-\ref{num:sec} below).


\section{Numerical Results} \label{num:sec}

Linear stability analysis of the solar wind plasma parameters given in Table~~\ref{model_param:tab}  are shown in Figure~\ref{fig:damping-rates} of Appendix~\ref{lin:sec}, indicating that all the studies cases are subject to kinetic  instabilities driven by temperature anisotropy and/or ion relative drift. In this section we describe the numerical results obtained with the 2.5D hybrid model (see, Appendix~\ref{model:sec}) of the proton and $\alpha$ core-beam populations for the model initial parameters given in Table~\ref{model_param:tab}, discuss and contrast the various cases of parameters sets, that demonstrate the effects of temperatures, temperature anisotropies and the beam speed  on the growth and relaxation of the ion kinetic instabilities, along with their associated ion-scale kinetic wave spectra in the nonlinear regime. 

\begin{table}[ht]
\caption{The dimensionless parameters of the numerical hybrid model runs for the various cases in the present study. The initial beam-core drift speed  $V_{d,p}$ for protons, the initial beam-core drift speed for $\alpha$ particles $V_{d,\alpha}$, the initial proton core density $n_{p,c}$, the initial proton beam density $n_{p,b}$, the initial $\alpha$ core density $n_{\alpha,c}$, the initial $\alpha$ beam density $n_{\alpha,b}$, the $\beta_\parallel$ of the ion populations (defined in terms of $n_e$), and the initial ion temperature anisotropies $A$.}

\hspace{-1.2in}
\begin{tabular}{ccccccccccccccc}
\hline
Case \# & $V_{d,p}$  & $V_{d,\alpha}$  & $n_{p,c}$  & $n_{p,b}$ &  $n_{\alpha ,c}$ & $n_{\alpha,b}$ & $\beta_{p,c}$ & $\beta_{p,b}$ & $\beta_{\alpha,c}$ & $\beta_{\alpha,b}$ & $A_{p,c}$ &  $A_{p,b}$ & $A_{\alpha,c}$ & $A_{\alpha,b}$ \\ \hline
	1  &  2            &  1.44       &  0.819 &  0.091 &  0.040 & 0.005 &  0.429 & 0.429 & 0.429 & 0.429 & 2 & 2 & 2 & 2\\
	2  &  2            &  1.44       &  0.819 &  0.091 &  0.040 & 0.005 &  0.214 & 0.858 & 0.214 & 0.852& 2 & 2 & 2 & 2\\
        3  &  2            &  1.44       &  0.819 &  0.091 &  0.040 & 0.005 &  0.214 & 0.858 & 0.214 & 0.852& 1 & 1 & 1 & 1\\
 	4  &  2            &    -          &  0.9    &  0.1     &  -         & -         &  0.214 & 0.858 & -      & -         & 2 & 2 & - & -  \\
	 5  &  2            &    -          &  0.9    &  0.1     &  -         & -         &  0.429 & 0.429 & -      & -         & 2 & 2 & - & -  \\
        6  &  2            &  2       &  0.819 &  0.091 &  0.040 & 0.005 &  0.429 & 0.429 & 0.429 & 0.429 & 2 & 2 & 2 & 2   \\
         7 &  1.4       &  2       &  0.819 &  0.091 &  0.040 & 0.005 &  0.214 & 0.858 & 0.214 & 0.852& 2 & 2 & 2 & 2

\end{tabular}
\label{model_param:tab}
\end{table}

In Figure~\ref{vxvz_a_p_pv2av1.4Ap2Aa2betac0.214betab0.858}a and b, we show the initial state of the proton and $\alpha$ particle populations VDFs in $V_x-V_z$ phase-space plane for the case with hot proton super-Alfv\'{e}nic beam ($V_{d,p}=2V_A$) and hot $\alpha$ particle super-Alfv\'{e}nic beam ($V_{d,\alpha}=1.44V_A$) with the beam temperatures about four times hotter than the core temperatures (see, Case~2 in Table~\ref{model_param:tab}). The temperature anisotropies are $A_p=A_\alpha=2$. This initial state is inspired by ion beam events deduced from SPAN-I observations, and the associated `hammerhead'-shaped velocity distributions \citep[e.g.,][]{Ver20,Ver22,McM24}. For reference, the initial states with isotropic temperatures are shown  in Figure~\ref{vxvz_a_p_pv2av1.4Ap2Aa2betac0.214betab0.858}e and f  (Case~3) with other parameters the same as in Case~2. The line plots show the $V_x$ dependence of the VDFs at $V_z=0$, demonstrating the Maxwellian core and the extent of the beam (tail) of the distribution.

\begin{figure}[h]
{\tiny\bf \hspace{0.3in} (a)\hspace{3.45in}(b)}\\
\includegraphics[width=0.5\linewidth]{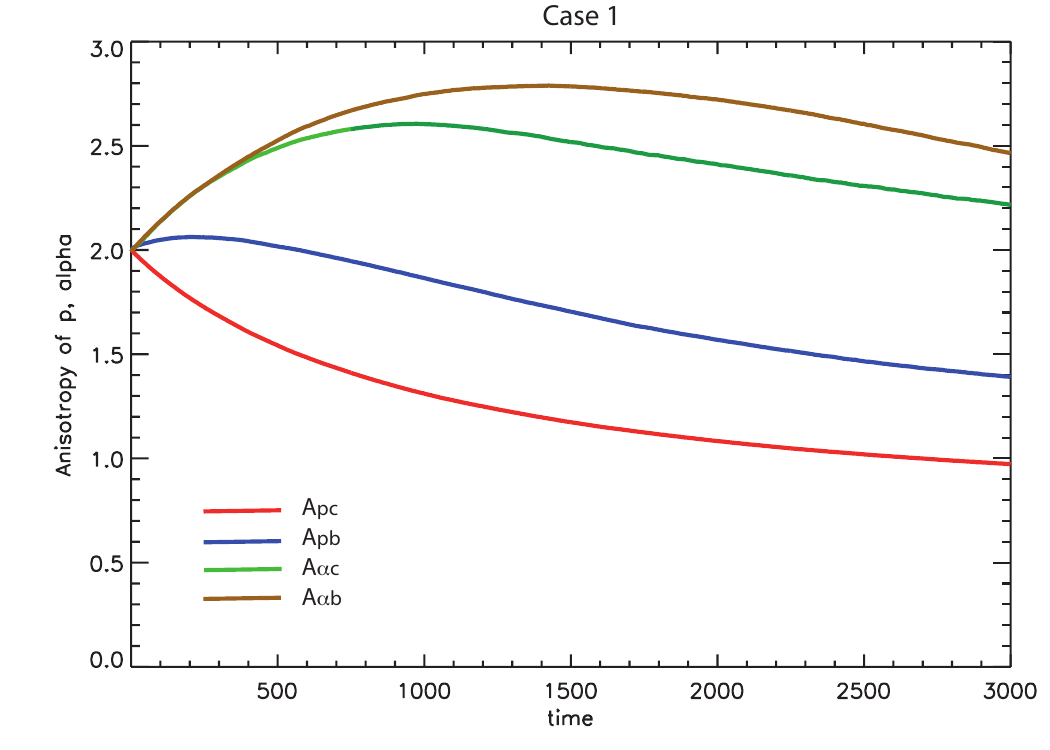}
\includegraphics[width=0.5\linewidth]{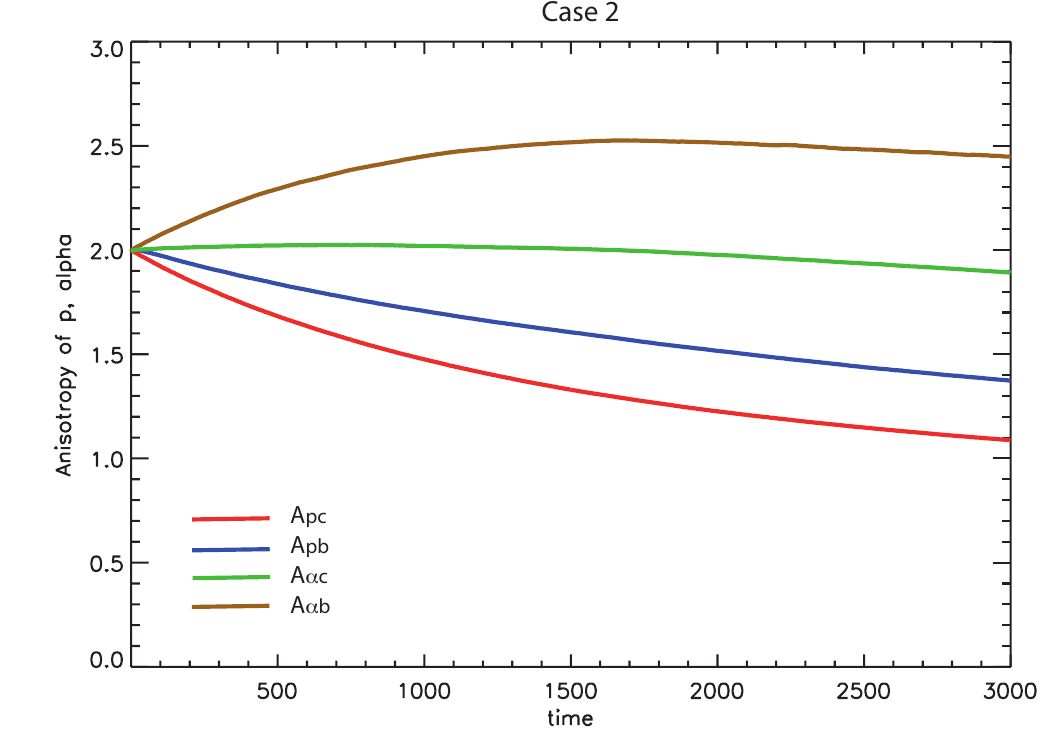}\\
{\tiny\bf\hspace{0.3in} .\ \ \ \  \ \ \ \  (c)\hspace{3.45in}(d)}\\
\includegraphics[width=0.5\linewidth]{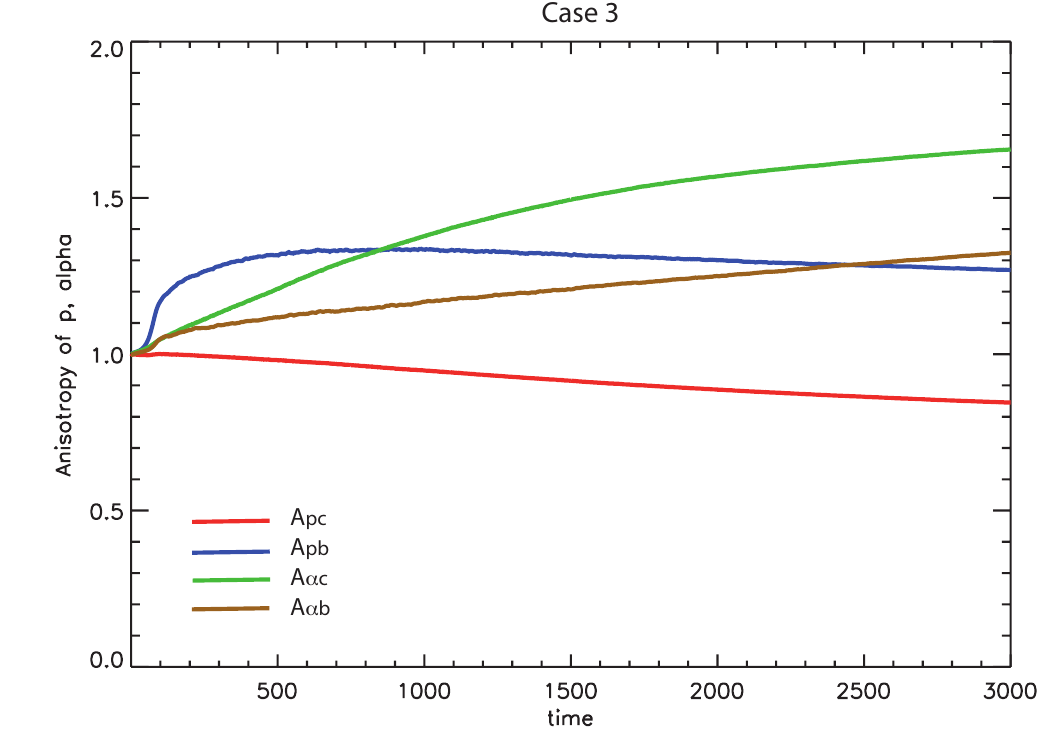}
\includegraphics[width=0.5\linewidth]{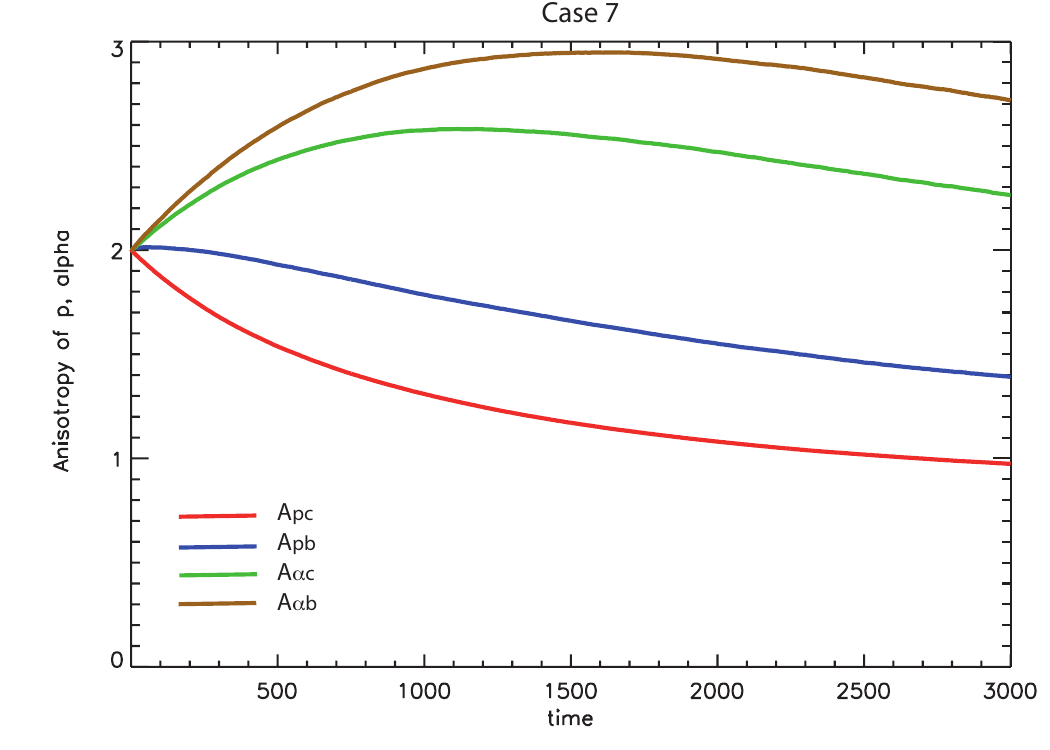}
\caption{The temporal evolution of the temperature anisotropies of the proton core (red), proton beam (blue), $\alpha$ particle core (green), and particle $\alpha$ beam (brown) for four cases in Table~\ref{model_param:tab}. (a) Case~1, (b)  Case~2, (c) Case~3 with  isotropic initial state, and (d) Case~7.}
\label{aniso_tfld:fig}
\end{figure}

The temporal evolution of the temperature anisotropies for $t=3000\Omega_p^{-1}$ of the proton and $\alpha$ populations for four cases in Table~1 (\#1, \#2, \#3, and \#7) are shown in Figure~\ref{aniso_tfld:fig}. The evolution of temperature anisotropies of the proton core, proton beam, $\alpha$ particle core, and $\alpha$ particle beam for Case~1 with initial parameters $V_d=2V_A$, $\beta_i=0.429$, $A_i=2$ for all populations are shown in Figure~\ref{aniso_tfld:fig}a. Figure~\ref{aniso_tfld:fig}b shows the temporal evolution of the temperature anisotropies for Case~2, where the ion beam populations are four times hotter than the cores. Figure~\ref{aniso_tfld:fig}c shows the temporal evolution of the temperature anisotropies for Case~3, that is same as Case~2 but with isotropic initial temperatures. Figure~\ref{aniso_tfld:fig}d shows the evolution for Case~7, that is same as Case~2 except the $V_{d,p}=1.4$, and $V_{d,\alpha}=2$. It is evident that in Case~1 the proton core temperature anisotropy decreases from the value of 2 at $t=0$, and becomes nearly isotropic after $t\approx 2500\Omega_p^{-1}$. In the same time interval, the proton beam anisotropy initially increases due to perpendicular heating by the ion-scale wave spectra produced by the magnetosonic drift instability, with subsequent relaxation to an anisotropy of $\sim1.4$ at the end of the run. Thus, the proton beam population can account for the anisotropic `hammerhead' part of the proton distribution observed by PSP/SPAN-I. The $\alpha$ particle population is heated in the perpendicular direction where both, core and beam population anisotropies increase to 2.6 and 2.8, respectively. The initial increase of the $\alpha$ particle temperature anisotropies is followed by subsequent relaxation of the anisotropies, due to the $\alpha$ population secondary instabilities. In Case~2 with initially cooler cores and hotter beams the evolution of the temperature anisotropies is more gradual compared to Case~1, with similar results at the end of the run. 

In Case~3 the initial state is the same as in Case~2, except that the initial temperatures of the four populations is isotropic. This leads to significantly different evolution compared to Case~2, where the proton beam and the $\alpha$ particle population are heated in the perpendicular direction. The proton beam reaching an anisotropy of $\sim1.3$ in about $600\Omega_p^{-1}$ following slight gradual decrease, while the  $\alpha$ core and beam populations temperature anisotropies increase continuously throughout the evolution, reaching anisotropies of 1.32 and 1.65, respectively at the end of the run ($t=3000$). The evolution of the temperature anisotropy in Case~7 in Figure~\ref{aniso_tfld:fig}d is very similar to Case~1 in Figure~\ref{aniso_tfld:fig}a, even though the beam drift of the protons and $\alpha$ particles are significantly different. This indicates that the anisotropy is not strongly affected by the super-Alfv\'{e}nic ion drifts at the modeled timescale but is mostly due to the initial anisotropies and the $\beta_\parallel,i$ of the ion populations.

\begin{figure}[h]
\centering
\includegraphics[width=0.8\linewidth]{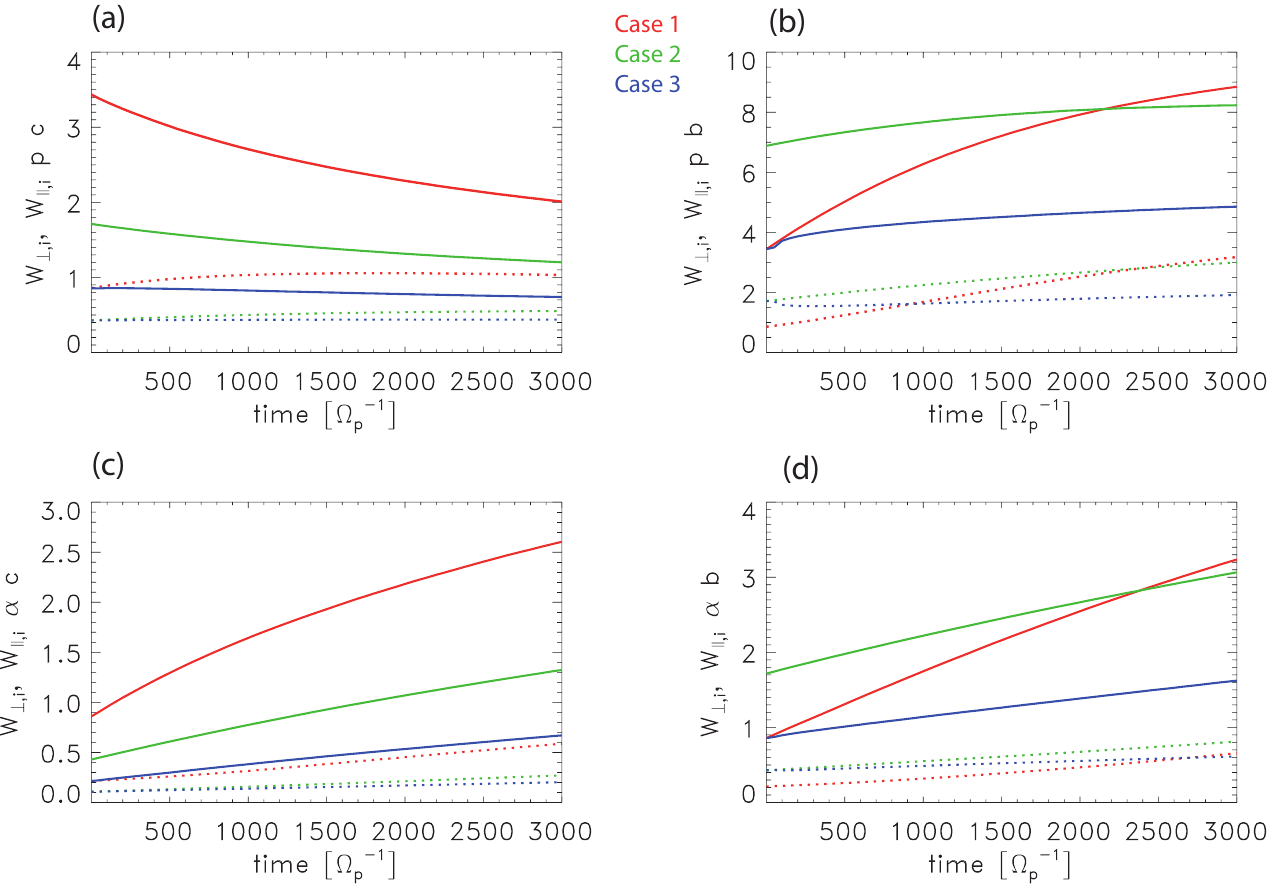}
\caption{The temporal evolution of the parallel and perpendicular thermal energies  of the proton core (a), proton beam (b), $\alpha$ particle core (c), and  $\alpha$ particle beam (d) populations  for Case~1 (red),  Case~2 (green), and Case~3 (blue). The line styles are $W_{\perp}$ solid, $W_{\parallel}$ dots. }
\label{wp:fig}
\end{figure}

The temporal evolution of the ion parallel and perpendicular thermal energies, $W_{\parallel,i}$, $W_{\perp,i}$ are shown in Figure~\ref{wp:fig} for Cases 1-3,  corresponding to the temperature anisotropies evolution shown in Figure~\ref{aniso_tfld:fig} for the duration $0-3000\Omega_p^{-1}$. The panels demonstrate how the ion heating is proceeding in the parallel and perpendicular directions with respect to the background magnetic field throughout the evolution and relaxation of the kinetic instabilities. It is interesting to note that in all initially anisotropic cases, the proton core population undergoes perpendicular energy decrease (cooling) due to the ion-cyclotron (IC) instability, thus providing energy to drive the ion-scale kinetic waves spectrum (see below), with additional contribution to the waves produced by the super-Alfv\'{e}nic ion beam-driven magnetosonic (MS) instability (see, Figure~\ref{Vd_t:fig} below). Note, that the hybrid model allows following the evolution of the initially core and beam particle populations separately, demonstrating the effects of the instabilities and the velocity-space diffusion in each of these population. The relaxation of the two (IC/MS) instabilities acting in parallel results in heating of the beam proton population, as well as the core and beam minority $\alpha$ particle population. It is evident that the proton beam is gaining energy in both directions, while in the $\alpha$ particles most heating is taking place in the perpendicular direction for both, core and beam populations. Even in the case of an isotropic initial state (Case~3) the evolution is qualitatively similar. However, in this case, the source of free energy comes entirely from the super-Alfv\'{e}nic beams, resulting in lower heating rate for the various ion populations, evolving to higher temperature anisotropies in the proton and $\alpha$ particle cores compared to the beams, contrary to the initially anisotropic core and beam cases.

In Figure~\ref{Vd_t:fig}, the temporal evolution of the proton and $\alpha$ particle beam-core velocity drift $V_d$ is shown for Cases~1, 2, 3. The decrease of $V_d$ takes place as a result of the wave-particle scattering of the beam ions toward the core as the beam instability is relaxed by emitting a spectrum of ion-scale kinetic waves. It is evident that in the anisotropic proton beam and core Case~1 and 2 the relaxation takes place at similar rates decreasing from $2V_A$ at $t=0$ to $1.6V_A$ and $1.5V_A$ in $3000\Omega_p^{-1}$,  respectively. In Case~3 with isotropic initial proton core and beam VDFs, the decrease in $V_d$ proceeds at faster rate, decreasing to $1.25V_A$ at $3000\Omega_p^{-1}$. The relaxation of the $\alpha$ particle beam velocity is slower than protons' due to the initially smaller drift of 1.4$V_A$, resulting in lower growth rate. This difference in relaxation rates can possibly be understood in terms of the contribution of wave spectra produced by the proton anisotropy-driven IC instability, in addition to the MS  instability in Cases~1 and 2. Thus, the combined effect of the IC/MS instabilities in the same ion population may result in slowing down the relaxation of the beam speed in comparison to Case~3, where the initial drifting proton VDF is an isotropic Maxwellian and there is only a single temperature anisotropy-driven IC instability in  each of the ion populations. The small difference in the relaxation rate between Cases~1 and 2 is consistent with the difference in the proton core and beam temperatures, where the initial beam temperature in Case~2 is twice as hot as the initial beam temperature in Case~1, while the initial core temperature in Case 2 is about half the initial core temperature in Case~1 (see, Table~\ref{model_param:tab}).
\begin{figure}[h]
\centering
\includegraphics[width=0.6\linewidth,height=0.35\linewidth]{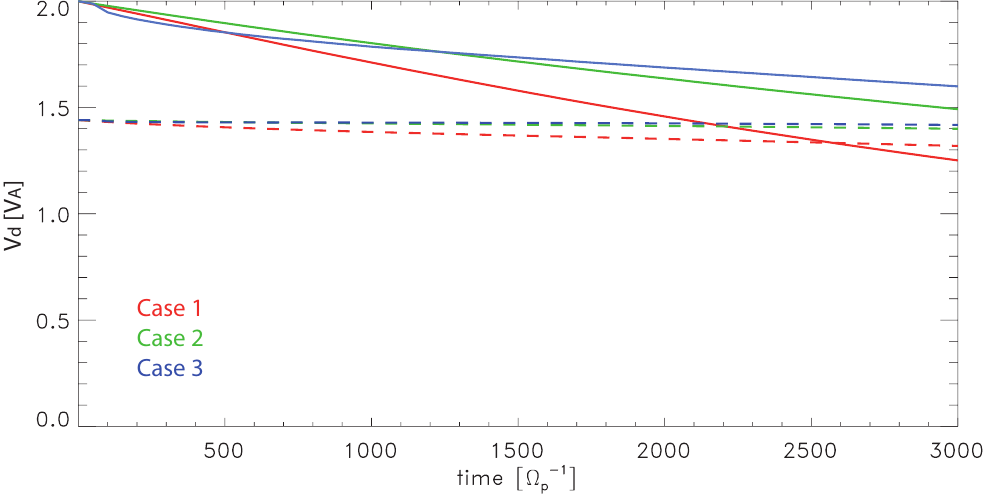}
\caption{The temporal evolution of the proton (solid) and $\alpha$ particle (dashed) beam-core drift velocities ($V_d$) for Case~1 (red), Case~2 (green), Case~3 (blue).}
\label{Vd_t:fig}
\end{figure}

The VDFs of protons and $\alpha$ particles in $V_x$-$V_z$ phase-space plane obtained from the 2.5D hybrid model at the end of the run at $t=3000\Omega_p^{-1}$ are shown in Figure~\ref{vxvz_a_p_pv2av1.4Ap2Aa2betac0.214betab0.858} for Case~2 (Figures~\ref{vxvz_a_p_pv2av1.4Ap2Aa2betac0.214betab0.858}a-d) and Case~3  (Figures~\ref{vxvz_a_p_pv2av1.4Ap2Aa2betac0.214betab0.858}e-h). The VDFs for the initial states at $t=0$ are shown in Figure~\ref{vxvz_a_p_pv2av1.4Ap2Aa2betac0.214betab0.858}a-b and e-f. It is evident in Case~3 that the core proton VDF has approached the anisotropic state through wave-particle interactions, while the beam speed has decreased to $V_{d,p}=1.5$, as evident from the 1D VDF cut along $V_x$, consistent with the value in Figure~\ref{Vd_t:fig} at $t=3000$. However, the proton beam VDF is still anisotropic and appears as a halo distribution, consistent with the 'hammerhead' proton VDF population, obtained from PSP/SPAN-I at perihelia \citep[e.g.,][]{Ver22}. The $\alpha$ particle population at the end of the run still exhibits strong perpendicular temperature anisotropy, in both, core and beam population, consistent with $\alpha$ particle VDFs obtained from PSP/SPAN-I \citep[e.g.,][]{McM24}. For initially isotropic Case~3 the proton core population temperature anisotropy decreases to $T_\perp/T_\parallel=0.85$, while the proton beam temperature anisotropy is $\sim1.27$, and the beam halo is apparent. The $\alpha$ particle population core and beam velocity distributions are anisotropic at the end of the run in Case~3, qualitatively similar to Case~2. The main differences are the somewhat smaller final values of the $\alpha$ particle beam velocity, as also evident in Figure~\ref{Vd_t:fig}. In the initially isotropic Case~3, the initial kinetic instability is entirely due to the super-Alfv\'{e}nic ion beams, and the somewhat lower free energy results in a more relaxed final state compared to the anisotropic Case~2. Thus, based on the 2.5D hybrid model, the study provides constraints on the required initial ion VDFs to achieve the final state consistent with observations. 
\begin{figure}[h]
\centering
(a)\hspace{1.5in}(b)\hspace{1.5in}(c)\hspace{1.5in}(d)\\
\includegraphics[width=0.24\linewidth]{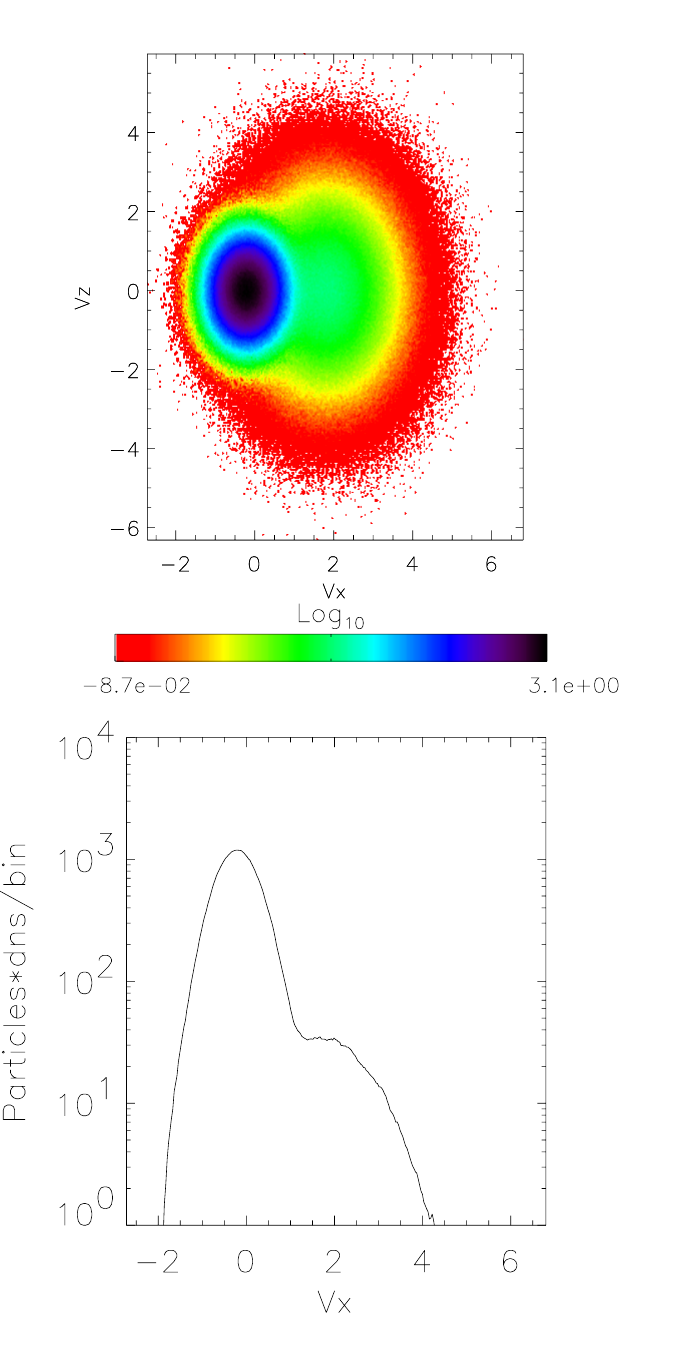}
\includegraphics[width=0.24\linewidth]{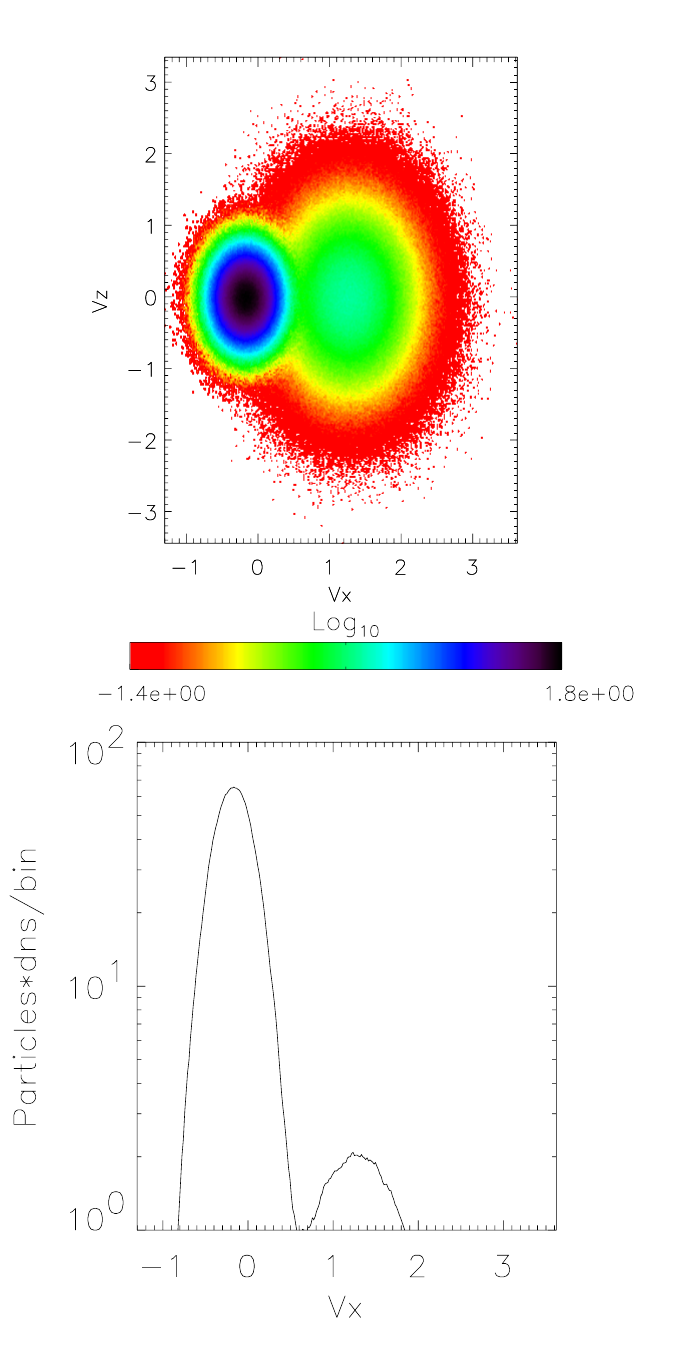}
\includegraphics[width=0.24\linewidth]{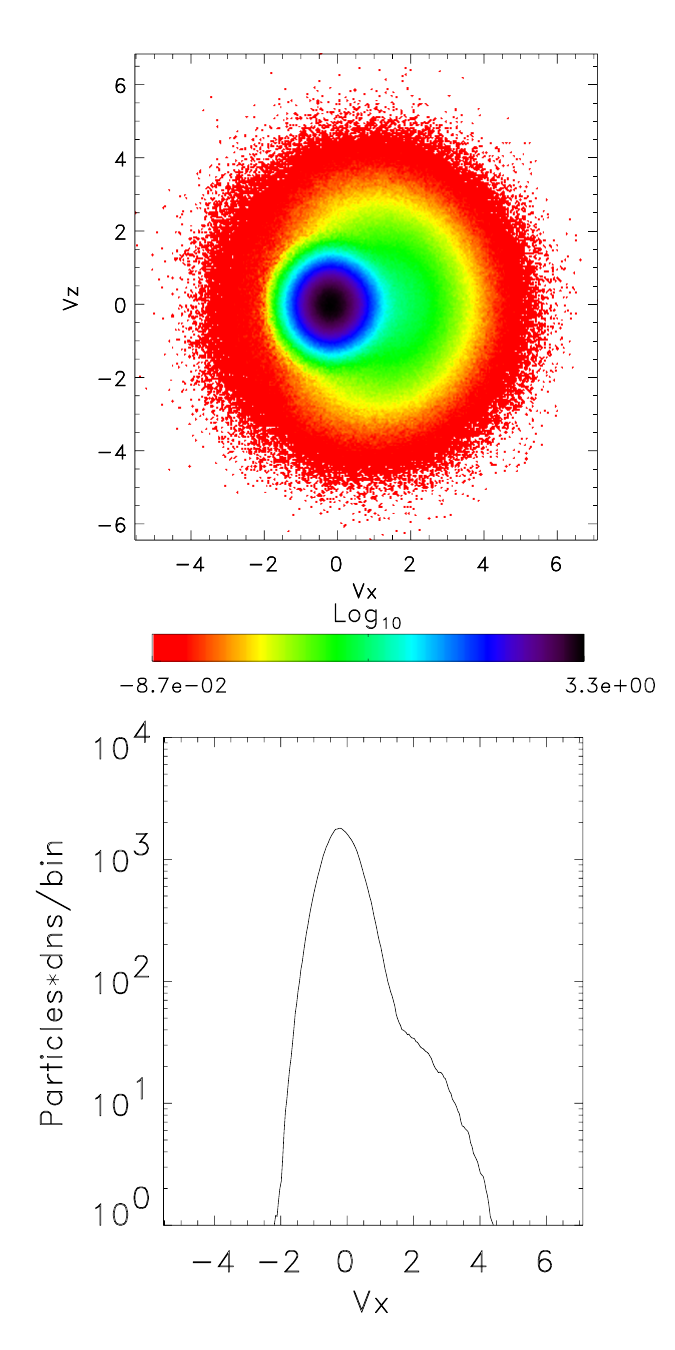}
\includegraphics[width=0.24\linewidth]{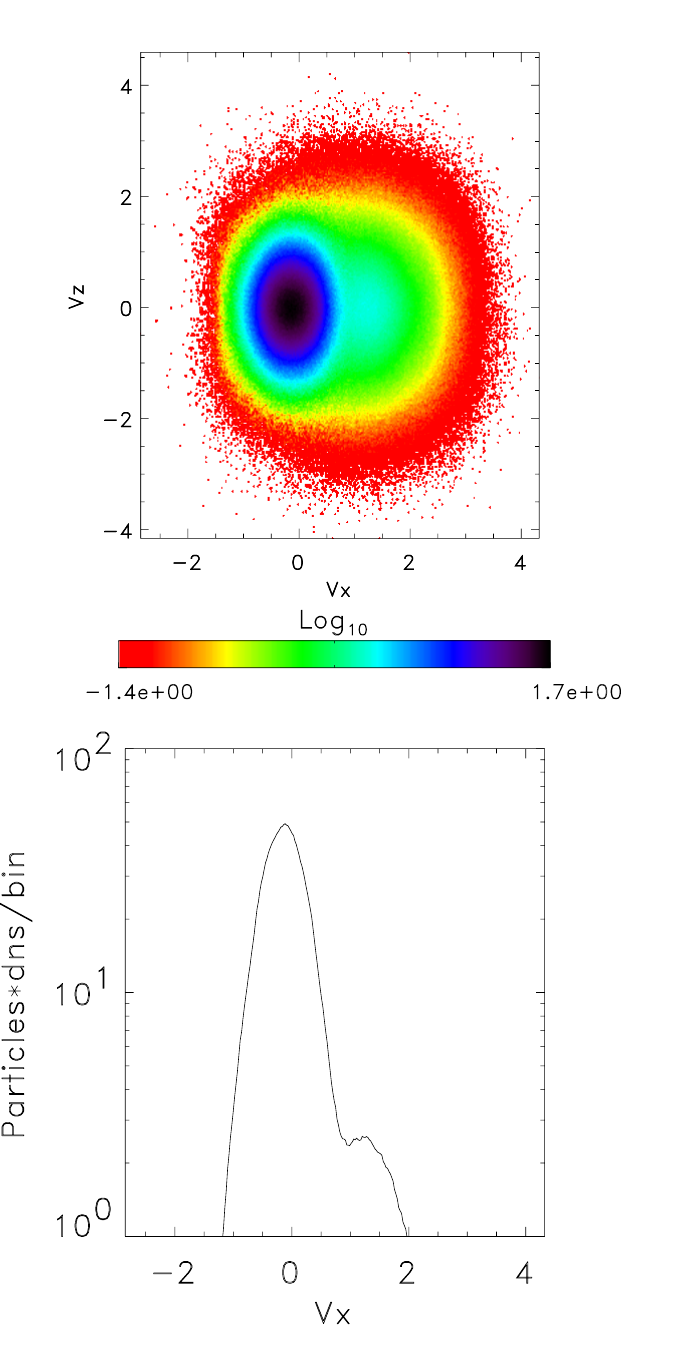}\\
(e)\hspace{1.5in}(f)\hspace{1.5in}(g)\hspace{1.5in}(h)\\
\includegraphics[width=0.24\linewidth]{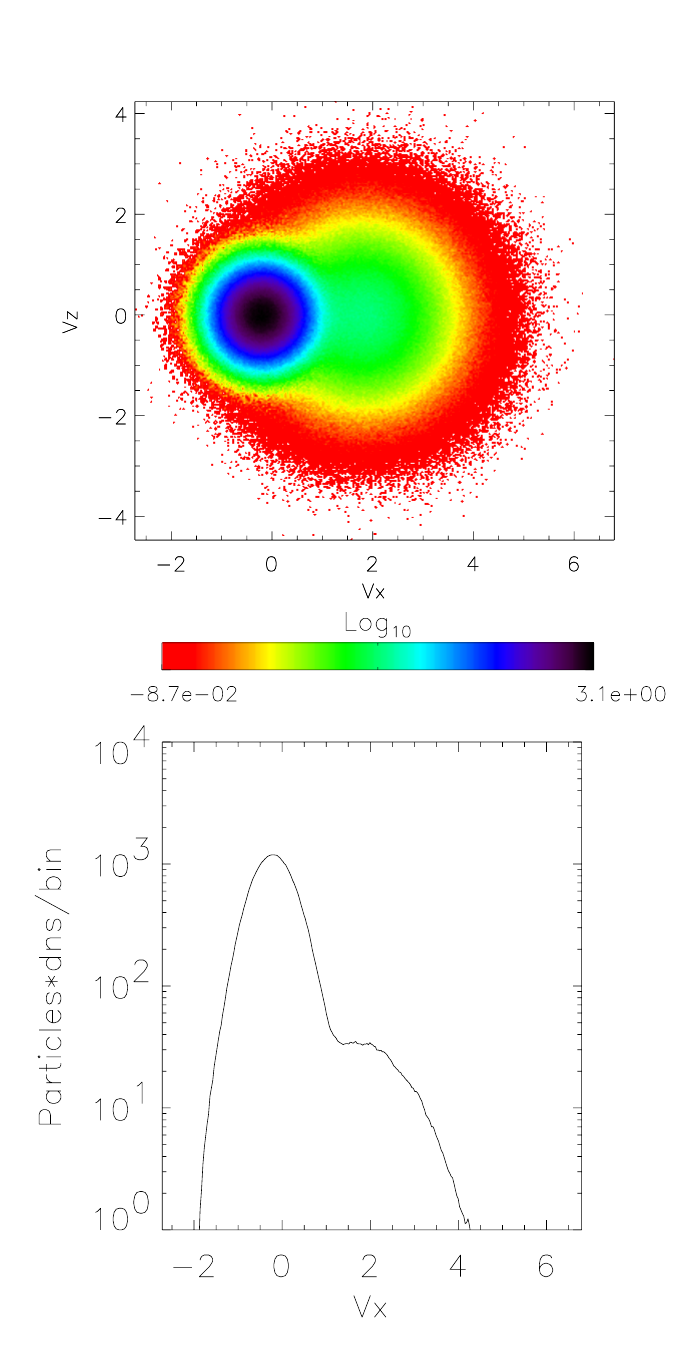}
\includegraphics[width=0.24\linewidth]{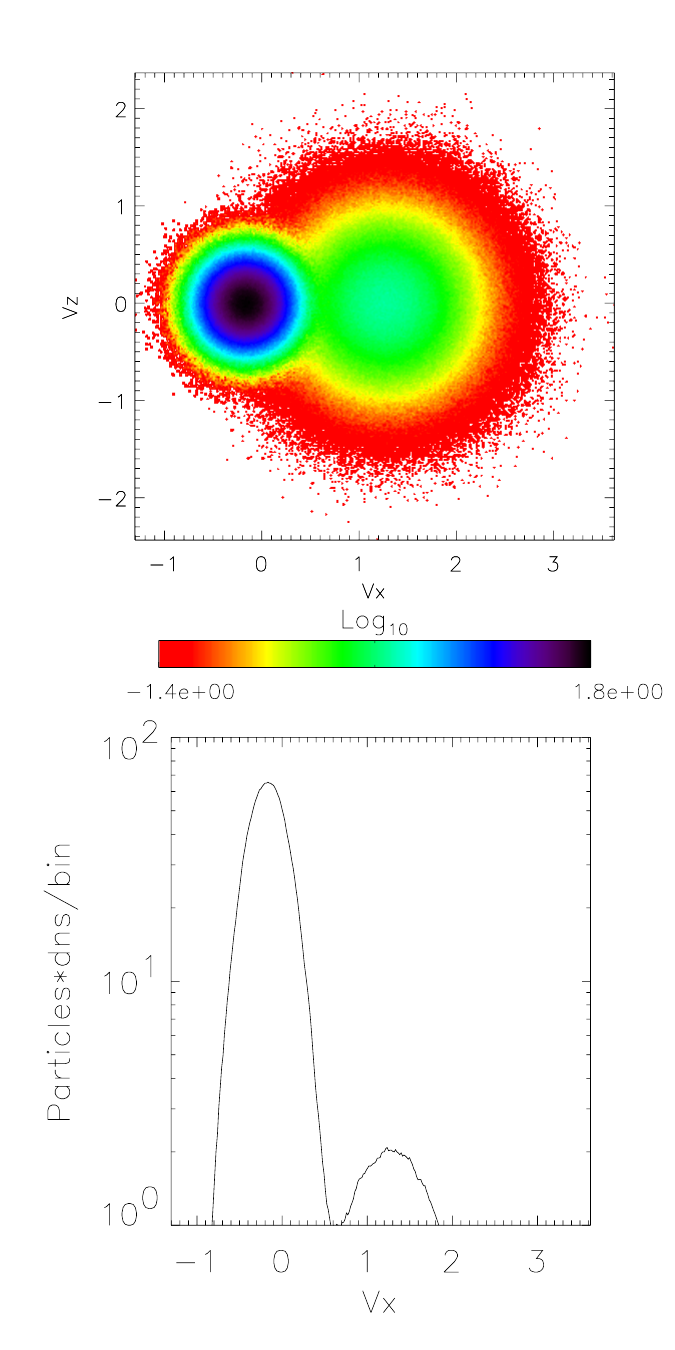}
\includegraphics[width=0.24\linewidth]{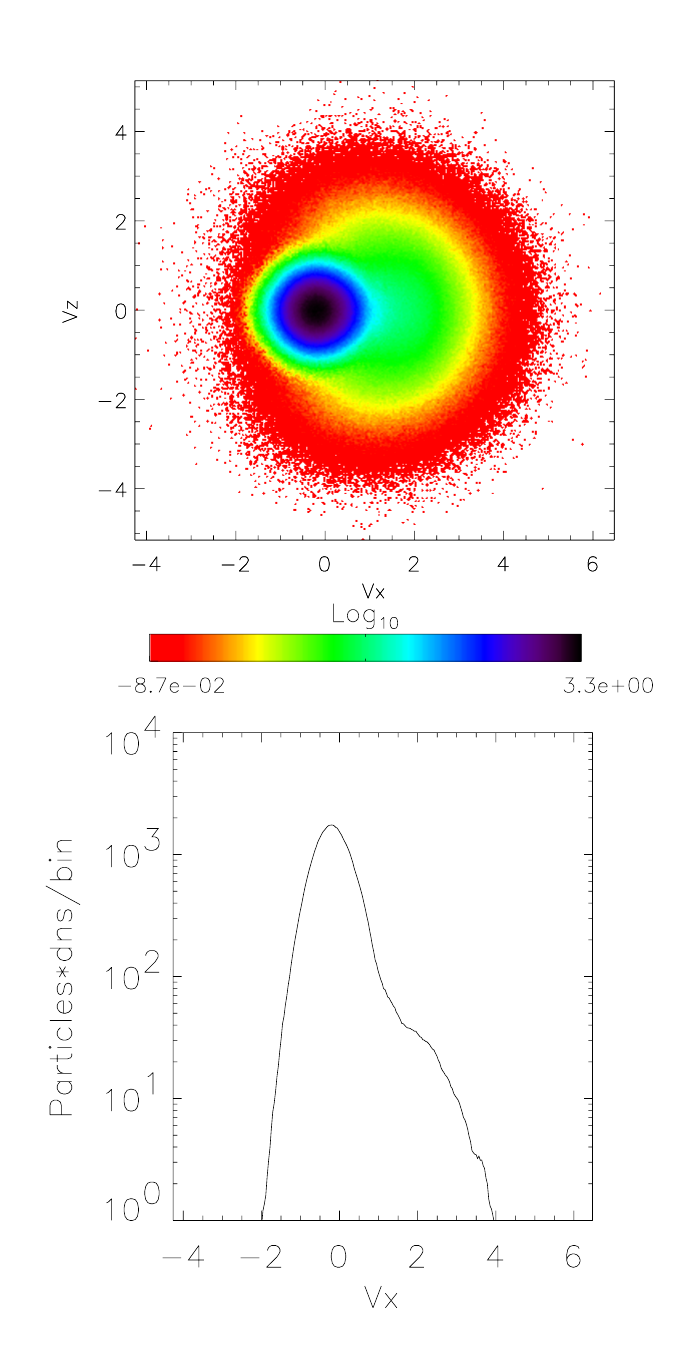}
\includegraphics[width=0.24\linewidth]{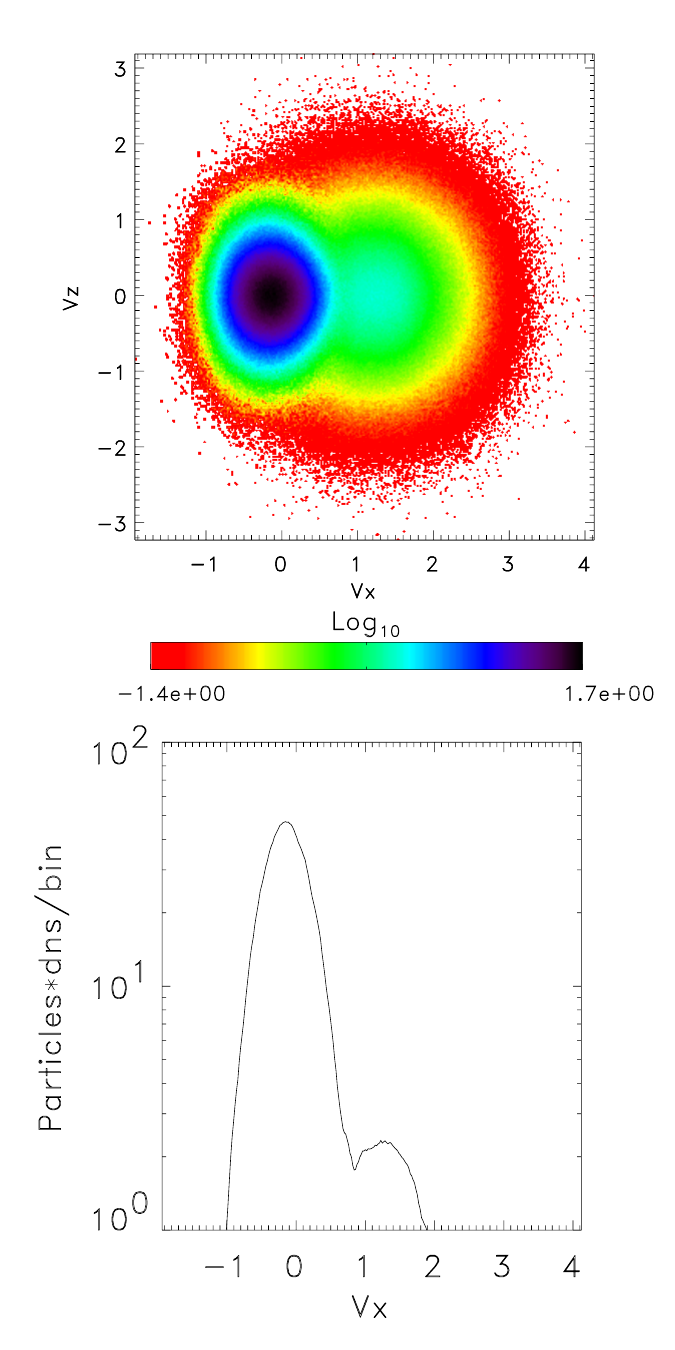}
\caption{The initial state of the VDFs and the results of the 2.5 D hybrid model at the end of the run for Case~2 (a) protons at $t=0$, (b) $\alpha$ particles at $t=0$, (c) protons at $t=3000\Omega_p^{-1}$, (d) $\alpha$ particles at $t=3000$. Same for Case~3 (e) protons at $t=0$, (f) $\alpha$ particles at $t=0$, (g) protons at $t=3000$, (h) $\alpha$ particles at $t=3000$. The corresponding line plots show the 1D VDFs dependence on $V_x$ at $V_z=0$. }
\label{vxvz_a_p_pv2av1.4Ap2Aa2betac0.214betab0.858}
\end{figure}

\begin{figure}[h]
\centering
(a)\hspace{2in}(b)\\
\includegraphics[width=0.3\linewidth]{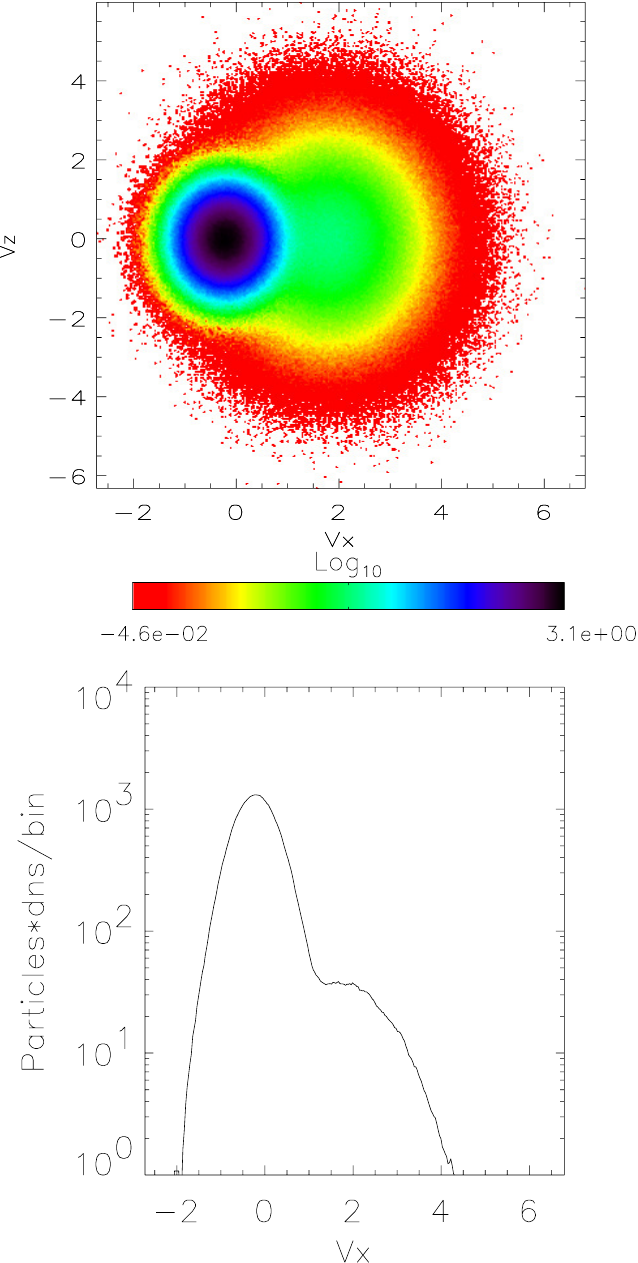}
\includegraphics[width=0.31\linewidth]{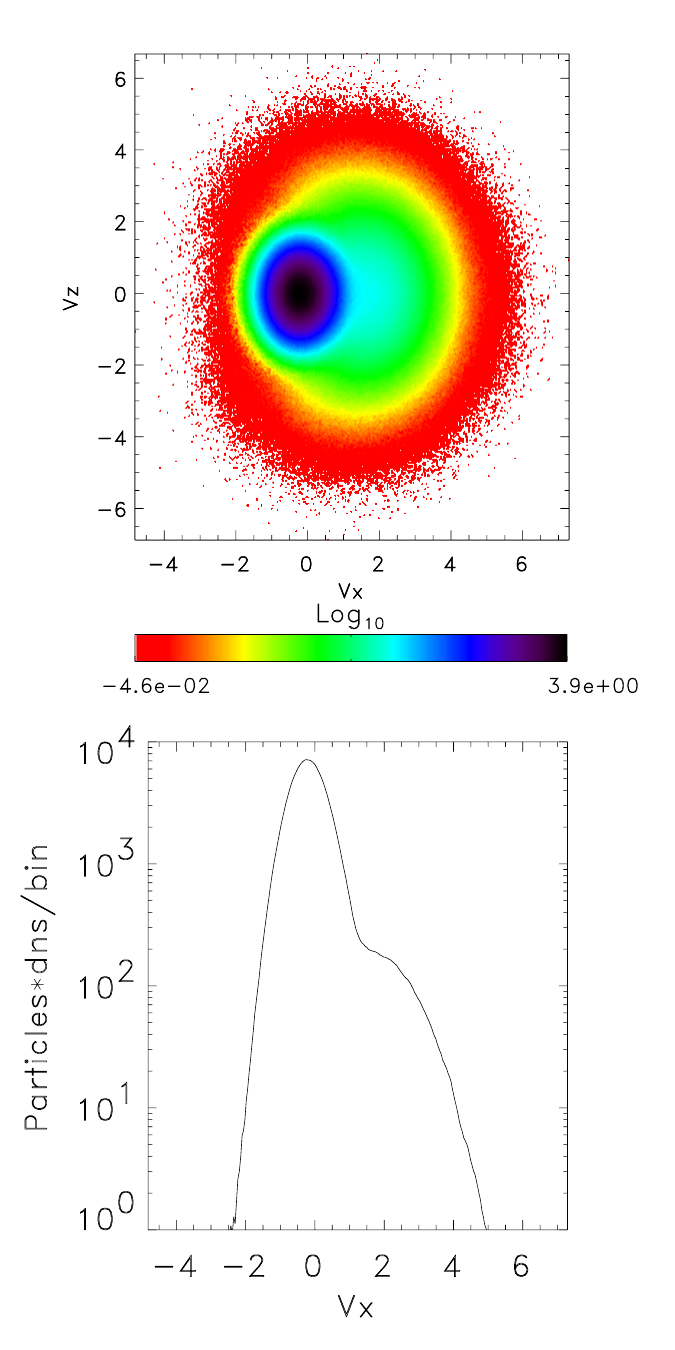}
\caption{ (a) The initial state of the proton core and beam VDFs in the 2.5D hybrid model for Case~4 (no $\alpha$ particles). (b)The evolved state of the proton VDFs at $t=600\tau_A$. The corresponding line plots show the 1D VDFs dependence on $V_x$ at $V_z=0$ (lower panels).  }
\label{vxvz_pv2Ap2Apb2betac0.214betab0.858_t0_t600:fig}
\end{figure}

For reference,  Figure~\ref{vxvz_pv2Ap2Apb2betac0.214betab0.858_t0_t600:fig}  shows the results of the run with only a proton population  (without $\alpha$ particles), where the proton beam speed is $2V_A$ and temperature anisotropy is 2 for proton beam and core populations, for the parameters in Case~4 in Table~1. The results show that the proton VDF evolution leads to perpendicular heating of the beam population, as well as the decrease of the drift speed $V_{d,p}$ are qualitatively similar to the previous cases with unstable $\alpha$ particle beams. Thus, the results demonstrate that the evolution of proton-related kinetic instability is not strongly affected by the presence of $\alpha$ particles.

\begin{figure}[h]
\centering
\includegraphics[width=0.6\linewidth,angle=90]{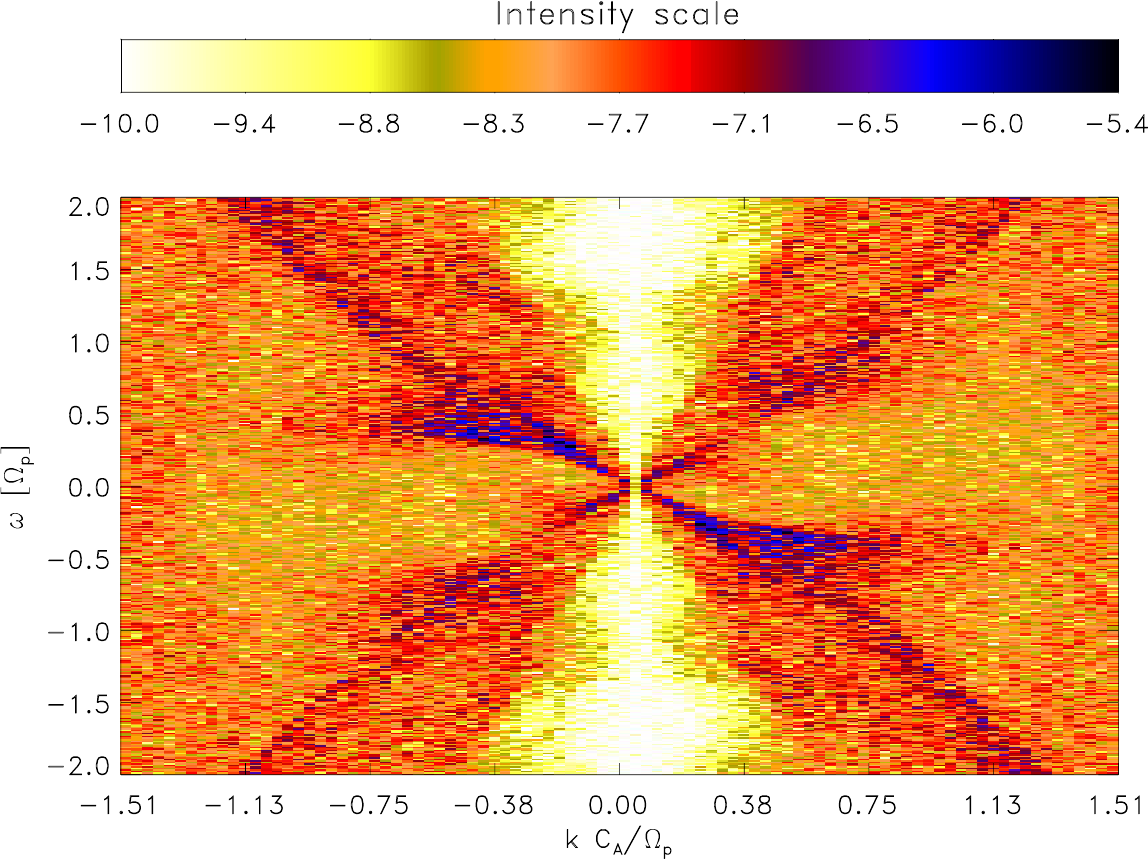}
\caption{The dispersion relation obtained from the 2.5D hybrid modeling results for Case~2. The frequency $\omega$ is in the unit of $\Omega_p$, while the $k$-vector is in the unit of inverse proton inertial length $\omega_{pp}/c=\Omega_p/V_A$, where $V_A$ is the Alfv\'{e}n speed, and $\omega_{pp}$ is the proton plasma frequency.}
\label{disp_pv2Ap2Apb2betac0.214betab0.858:fig}
\end{figure}

The dispersion relation obtained from the 2.5D hybrid code for Case~2 is shown in Figure~\ref{disp_pv2Ap2Apb2betac0.214betab0.858:fig} throughout the evolution of the beam instabilities. The power in the various dispersion branches that can be identified from linear Vlasov's dispersion \citep[see, e.g.,][]{MOV15} is evident, where the nonresonant branches extend to higher frequencies and higher $k$ magnitude nearly undamped. The resonant branches contain significantly higher power (shown in blue with $log_{10}$ magnitude scale). These branches dissipate at $k>0.7$,  at frequencies in the range $0.5\Omega_p<\omega< 1\Omega_p$, consistent with proton and $\alpha$ particle resonance frequencies. Note that due to the Doppler shift caused by the beams the dispersion branches are not symmetric with respect to the axis, and the Doppler shift affects the resonant condition for the ions such that $\omega_i-k_\parallel V_\parallel=\Omega_i$, where the parallel direction is along the magnetic field in the field-aligned beams \citep[see, e.g.,][]{XOV04} (to clarify, this modeled Doppler shift is unrelated to the observational Doppler shift, discussed in Section~\ref{obs:sec} above).

\begin{figure}[h]
\centering
\includegraphics[width=0.35\linewidth]{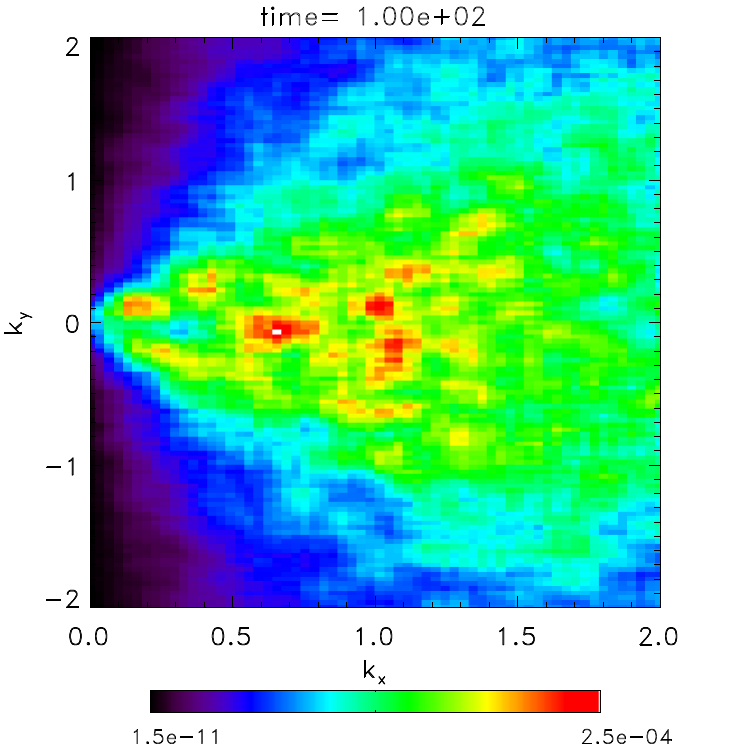}
\includegraphics[width=0.35\linewidth]{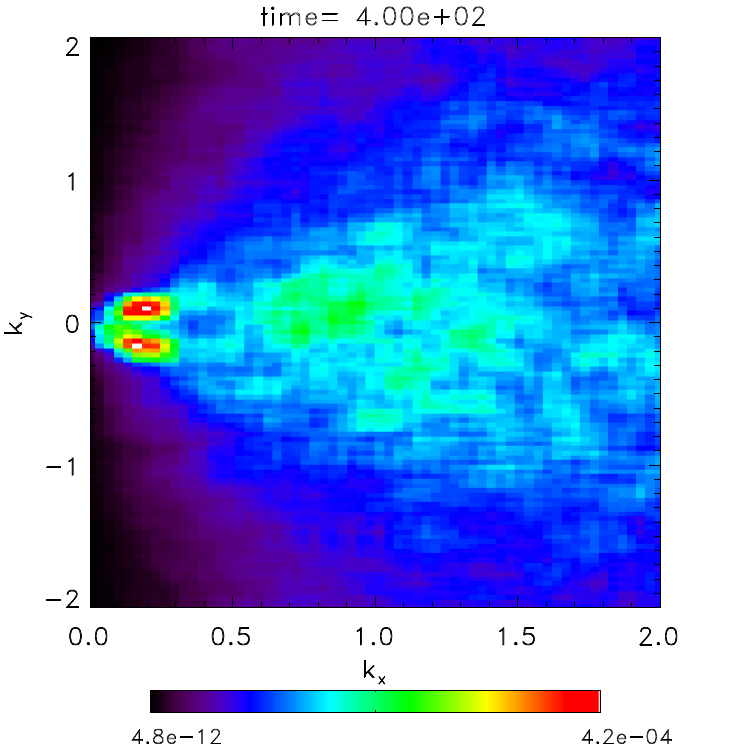}
\includegraphics[width=0.35\linewidth]{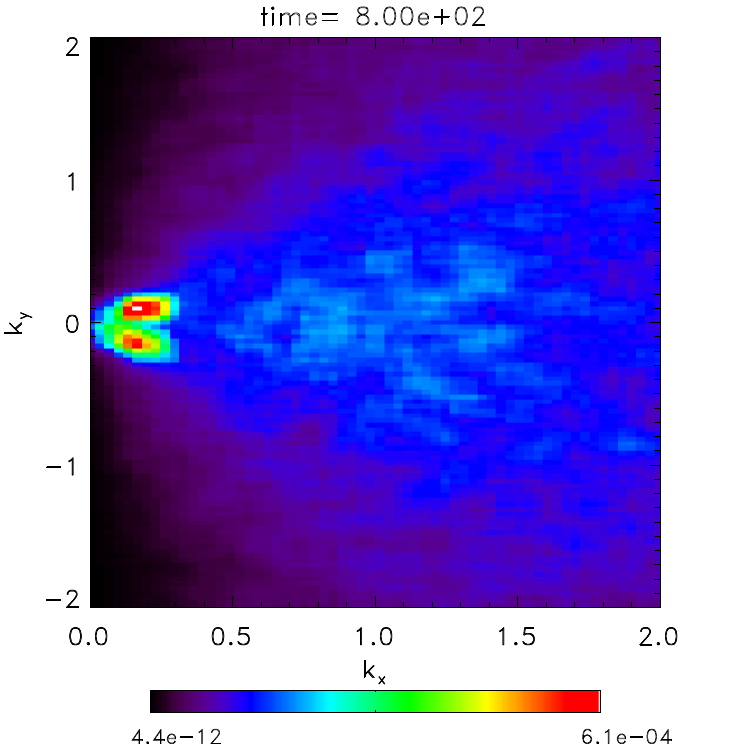}
\includegraphics[width=0.35\linewidth]{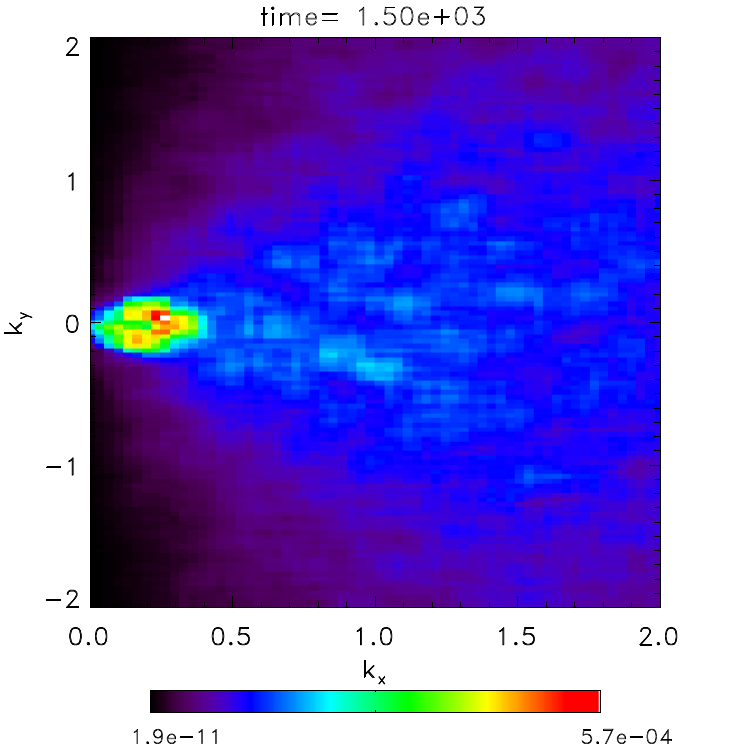}
\caption{The $k_x-k_y$ power spectrum of the perpendicular magnetic field component ($B_y$) obtained from the 2.5D hybrid modeling at $t=100,\ 400,\ 800,\ 1500\,\Omega_p^{-1}$ for Case~2. The $k$-vector is in units of inverse proton inertial length $\omega_{pp}/c=\Omega_p/C_A$.}
\label{2dfft_kxky_case2:fig}
\end{figure}

In Figure~\ref{2dfft_kxky_case2:fig}, the snapshots of the $k$-spectrum of  magnetic fluctuations in the $k_x - k_y$ plane are shown for Case~2 at times $t=100,\ 400,\ 800,\ 1500\,\Omega_p^{-1}$.  It is evident that at the beginning of the evolution, there is a significant power in oblique modes ($|k_y|>0$) with most peaks of power at $k_x>0.5$. However, after initial growth, the power at higher values of $k_x$ has dissipated with the peaks moving to lower values of $k_x<0.15$. It is interesting to note that in the nonlinear stage of the evolution, the $k$-power spectrum does not evolve significantly compared to the rapid change in the initial state. Also, the power remains in somewhat oblique modes in this case. The evolution is typical of the beam-driven ion-scale wave spectrum, qualitatively similar to other cases of this study.


\section{Discussion and Conclusions} \label{disc:sec}
Comprehensive evidence of complex anisotropic non-Maxwellian proton velocity distributions comprising core, beam, and `hammerhead' populations has been extensively found in recent observations of the solar wind ions by the SPAN-I instruments on board the PSP spacecraft from Encounter 4 onward. The beams have near- or super-Alfv\'{e}nic speed in relation to the core, and both populations show temperature anisotropies with $T_\perp/T_\parallel>1$. Kinetic ion-scale waves were detected by the FIELDS instrument with both left-hand (LH) and right-hand (RH) polarization,  linked to the ion beams. In the present study we use data from Encounter 17 to demonstrate the observation of proton and $\alpha$ particle VDFs with super-Alfv\'{e}nic beams and associated ion-scale kinetic waves. We find strong evidence of ICWs that are likely produced by the IC instability as well as evidence of both left-hand and right-hand polarized waves that could be associated with the beam instability.  

Motivated by recent PSP observations at perihelia, we use linear stability analysis, followed by nonlinear hybrid models to study the evolution of beam and ion temperature anisotropy-driven instabilities produced by hot anisotropic proton and $\alpha$ particle beams. We find that the instabilities saturate nonlinearly and lead to local plasma heating, converting the free energy in the unstable distributions to random (thermal) solar wind plasma energy, thus leading to solar wind heating. The modeling show that parallel propagating ICWs as well as oblique and magntosonic waves are produced. We find that the beam population of the ions exhibits the `hammerhead' structure in qualitative agreement with the PSP observational findings of ions VDFs at perihelia, as well with the observed ion-scale wave activity. Our modeling results support the understanding that collisionless waves-particle interactions provide an important step in converting the large scale magnetic fluctuations energy to thermal energy and anisotropic heating of the solar wind plasma. 

It is interesting to compare the results of the evolution of the temperature anisotropies (Figure~\ref{aniso_tfld:fig}) and the ion beam velocities (Figure~\ref{Vd_t:fig})  in the present study to the previous cold isotropic ion beam results of \citet{Ofm22a} (see their Figure~7). It is evident that in the present  hot core and beam ion plasma the anisotropies and the beam velocities of the proton and $\alpha$ particle populations evolve more gradually and the anisotropies increase to lower peak values, compared to the cold isotropic beam case studied in \citet{Ofm22a}. This could be understood in terms of lower instabilities growth rates, and the lower values of  free energy in the super-Alfv\'{e}nic drift relative to the hot beam thermal energy in the present model compared to the cold beam cases. We find that the present modeling results are in good qualitative agreement with the later perihelia observations, such as ion VDF morphology, `hammerhead' beams, and ion-scale kinetic wave activity.

While the particle velocity distributions and related instability processes presented in this work can contribute to the energization of the young solar wind, it is important to understand how frequently such velocity distributions in the young solar wind are presented, along with recognizing the situations when they correspond to the developing ion-kinetic instabilities. This type of analysis can be potentially tackled with the development of machine learning (ML) diagnostics techniques, with ML models targeted at automatically identifying unstable VDFs. Recently, \citet{Mart23} have developed an ML-driven linear instability detection and categorization model, Stability Analysis Vitalizing Instability Classification (SAVIC)\footnote{\url{https://savic.readthedocs.io/en/latest/}}, that allows one to efficiently categorize the linear stability of the VDFs consisting of bi-Maxwellian anisotropic distributions of the proton core, proton beam, and $\alpha$ particle core components. Currently, \citet{Sad25} are investigating the potential of expanding the ML diagnostics to the cases in the non-linear particle-wave interaction regimes using the dataset of Hybrid-PIC simulations. The Hybrid-PIC runs analyzed in this paper are included in the effort.

\begin{acknowledgments}
We thank the PSP mission team for generating the data and making them publicly available. The authors acknowledge support by NASA grants 80NSSC20K0648, 80NSSC24K0724. L.O., S.A.B., Y., and V.S. acknowledge the support by NSF grant AGS-2300961. 
M. M. M. and K. G. K. were financially supported by NASA grants:  80NSSC22K1011, 80NSSC19K1390, 80NSSC23K0693, 80NSSC19K0829, and 80NSSC24K0724.
P. M. acknowledges the partial support by NASA HGIO grant
80NSSC23K0419 and the NSF SHINE grant 2401162.
Resources supporting this work were provided by the NASA High-End Computing (HEC) Program through the NASA Advanced Supercomputing (NAS) Division at Ames Research Center. This study benefited from discussions at the International Space Science Institute (ISSI) in Bern, through ISSI International Team project 563 (Ion Kinetic Instabilities in the Solar Wind in Light of Parker Solar Probe and Solar Orbiter Observations) lead by L. Ofman, and L. K. Jian.
\end{acknowledgments}

%

\vspace{5mm}
\facilities{PSP (SWEAP, FIELDS)}





\appendix

\section{Linear Stability Analysis}\label{lin:sec}

The role of secondary $\alpha$ particles, or $\alpha$ beams, in generating right-handed (RH) polarized ion-scale waves was recently studied using linear stability analysis by \cite{McM24}, where they demonstrated the occurrence of RH and LH waves alongside the parameters of $\alpha$ and proton populations during the observed event. 
Combining plasma data from PSP and the predictions of linear instability from the \texttt{PLUME} dispersion solver \citep{Kle15}, the study provided insights into how $\alpha$ particles contribute to wave generation in the solar wind. 
For the interval studied, \cite{McM24} found that kinetic instabilities are the primary drivers of the ubiquitous waves seen in the inner heliosphere. 
These findings from linear analysis together with present nonlinear hybrid modeling results described below, as well as past studies \citep[e.g.][and references within]{Ofm22a}, emphasize the importance of $\alpha$ particles in energy transfer and wave-particle interactions, which are critical to understanding solar wind dynamics. 

\begin{figure}
    \centering
    \includegraphics[width=0.95\linewidth]{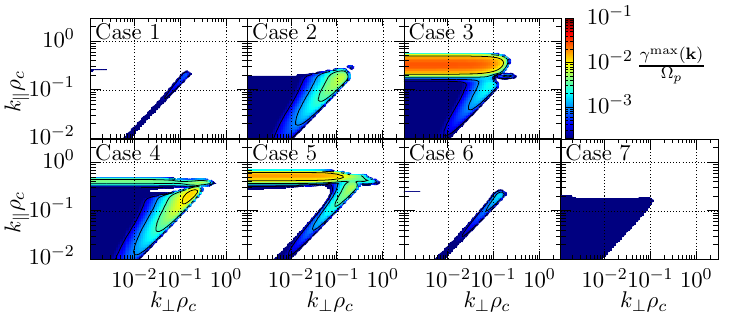}
    \caption{Linear growth rates as a function of wavevector $\gamma^{\textrm{max}}(\mathbf{k})/\Omega_p$ for the seven cases described in Table~\ref{model_param:tab}.
    All cases are linearly unstable, with the relative importance of parallel and oblique instabilities sensitively depending on the contributions between the ion temperatures and drifts.}
    \label{fig:max-gamma}
\end{figure}

To constrain the impact of these different ion populations, we run nonlinear simulations for seven cases of linearly unstable plasmas, outlined in Table~\ref{model_param:tab}.
All seven cases have electron, a proton core, and proton beam populations, with five of the seven having an $\alpha$ particle core and beam populations as well, and the remaining two serving as counter examples for when the $\alpha$ populations are not present.
Before running the nonlinear simulations, we consider the linear stability of these seven cases.
This is done by determining the growth rate of the most unstable modes at a given wavevector by evaluating the Nyquist criterion.
The linear calculation is performed using the \texttt{PLUMAGE} code \citep{Kle17}, which numerically performs a contour integral over the complex dispersion surface associated with the susceptibilities from the prescribed populations to determine the number of unstable modes.
This method has been successfully applied to spacecraft data as a means of determining the linearly stability of large ensembles of observations \citep{Kle18,Mar21b}.
Fig.~\ref{fig:max-gamma} illustrates the growth rate as a function of wavevector $\gamma^{\textrm{max}}(\mathbf{k})/\Omega_P$ for the seven cases.
We emphasize that this calculation only identifies the fastest growing mode at a particular wavevector, and that at some of these wavevectors there are other, more slowly growing, instabilities that may be present due to many sources of free energy associated with the complex velocity distributions.

All of the cases support linearly unstable solutions, though there is a significant difference in the maximum growth rates between them.
For these distributions, we see two classes of instabilities, those that a primarily parallel propagating and those that have oblique wavevectors.
By comparing Cases 1 and 2, we see that making the core colder and the beam hotter for both the protons and $\alpha$ particles, we significantly enhance the growth rate of the oblique instability.
Making all of the ion populations isotropic, but retaining the relative drifts, Case 3, leads to the generation of strongly growing parallel propagating mode.
Removing the $\alpha$ particle contributions altogether in Case 4, promotes the oblique instability to being the most unstable mode again. 
Returning the proton temperatures back to the case 1 values, but with no $\alpha$ particle contributions, leads to a much more unstable set of modes.
The differences between Cases 1 and 6, and Cases 2 and 7 illustrate the impact of the relative drifts between the core and beam populations, leading to orders of magnitude changes in the maximum growth rate.
\begin{figure}
    \centering
    \includegraphics[width=0.95\linewidth]{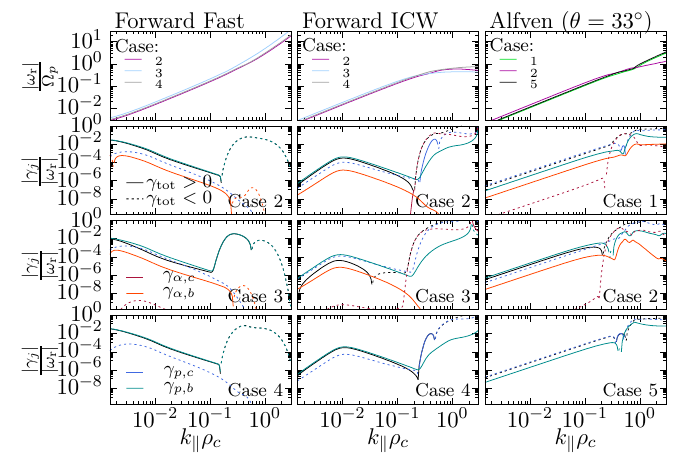}
    \caption{Linear frequencies and damping rates for the parallel forward fast (left column), parallel forward Alfv\'{e}n/ICW (center) and oblique Alfv\'{e}n solutions.
    (Top row) Real frequency $\omega_{\textrm{r}}/\Omega_p$ for the designated solution for three selected cases.
    (Second, Third and Fourth Rows) Power absorption and emission per wave period from population $j$, $|\gamma_j|/|\omega_{\textrm{r}}|$, as well as the total absorption or emission. 
    Solid and dashed lines represent positive and negative values respectively.
    The proton core and beam are shown in blue and green, with the $\alpha$ core and beam in red and orange.
    }
    \label{fig:damping-rates}
\end{figure}

To understand the couplings that drive these changes in the linear stability, we determine the linear dispersion relations for the seven cases using \texttt{PLUME}, including their complex frequencies as well as the power emitted or absorbed by each component $j$ \citep{Qua98,Kle20},
\begin{equation}
    \frac{\gamma_j}{\omega_{\textrm{r}}} = \frac{\mathbf{E}^* \cdot \underline{\underline{\chi}}_j^a \cdot \mathbf{E}}{4 W_{\textrm{EM}}},
\end{equation}
where $\mathbf{E}$ and $\mathbf{E}^*$ are the electric field eigenfuctions associated with the normal mode solution and its complex conjugate, $\underline{\underline{\chi}}_j^a$ is the anti-Hermitian part of the susceptibility tensor for component $j$ and $W_{\textrm{EM}}$ is the electromagnetic wave energy.
By looking at the amplitude and signs of these rates  we can assess the impact of individual components on the overall stability of a solution.
We show in Figure~\ref{fig:damping-rates} three different solutions across selected cases, specifically the forward parallel propagating fast/whistler solution, the forward parallel propagating Alfv\'{e}n/ICW solution, and the oblique Alfv\'{e}n solution, with an angle between $\mathbf{k}$ and the mean magnetic field of $33^\circ$.
We see that there is relatively little variation in the real frequency for a given mode between the cases.
However, the stability and the components that contribute to the overall damping vary significantly between the cases.

For the forward fast mode, we have a large scale, slowly growing instability driven by the drift in the proton beam population.
Reducing the temperature anisotropy of the beam, while keeping the same drift speed, slightly increases the phase speed of the mode and causes the beam to drive a faster growing instability at $k_\parallel \rho_c \sim 0.3$.
The $\alpha$ particle populations do not significantly impact the behavior of this solution, while they do alter the behavior of the forward Alfv\'{e}n wave.
The proton core is cyclotron unstable due to its temperature anisotropy, which can be seen by comparing Cases 2 and 3. 
However, the $\alpha$-core population is able to absorb the power emitted from the proton core, rendering the entire mode stable at scales above $k_\parallel \rho_c \sim 0.2$.
Removing the $\alpha$'s, case 4, allows the entire mode to become ICW unstable.

For the oblique Alfv\'{e}n solution, we see a sensitive dependence on the temperatures of the populations.
Allowing the beam populations to be four times hotter than the core populations, comparing Cases 1 and 2, the power emitted by the beams overtakes the power absorbed by the cores, leading to an unstable solution.

With the linear stability parameterized, we next turn to nonlinear hybrid simulations of these seven cases.

\section{Brief Hybrid-PIC Model Description} \label{model:sec}
Previous studies, such as \citet{Kle21,Ver22,McM24} used linear dispersion solvers to analyze stability and wave-particle interactions. However, the evolution of VDFs and ion-scale wave generation cannot be fully and self-consistently captured by linear methods. To explore these processes in more detail, we used a hybrid model, which provides a more realistic representation of energy transfer between particles and waves, including important nonlinear interaction. Below, we investigate the evolution of hot proton and $\alpha$ beams using the hybrid model, providing deeper insights into these phenomena and supporting future exploration of hot beams in data from recent PSP encounters. 

In this study, we employ the hybrid-PIC model (hereafter the hybrid model), with two ion species (protons and $\alpha$'s). The method of solution is the same as in our previous studies using the 2.5D and 3D hybrid codes, described in our previous papers \citep[e.g.,][and references therein]{OV07,Ofm10,OVM14,Ofm22a}. Here, we provide a brief description of the hybrid model for reference and convenience. The hybrid codes solves the equations of motions of the ions modeled as an ensemble of large number of Gaussian-shaped super-particles subject to the forces of the background electro-magnetic fields, assuming quasi-neutrality of the plasma ($n_p+2n_\alpha=n_e$, where $n_p$ is the proton number density, $n_\alpha$ is the $\alpha$ particle number density, and $n_e$ is the electron number density). A generalized Ohms' law is solved for the electric field by neglecting the electron mass, and Maxwells' equations are solved for $B$ on the finite grid of cells (typically $128^2$ or $256^2$) with 512 particles per cell. The boundary conditions are periodic. The solutions are advanced in time by using the Rational Runge-Kutta (RRK) method \citep{Wam78}. The velocities are normalized by the Alfv\'{e}n speed, $V_A$, the distances by the proton inertial length $\delta_p=c/\omega_{pp}$, where $\omega_{pp}$ is the proton plasma frequency), and the time is normalized by the inverse proton gyroresonat frequency $\Omega_p^{-1}$, where  $\Omega_p=eB/m_pc$, where $B$ is the background magnetic field magnitude, $m_p$ is the proton mass, and $c$ is the speed of light. The typical time step is $\Delta_t=0.01\Omega_p^{-1}$, and the spatial resolution is $0.5\delta$ in the two dimensions. The initial state of the plasma is spatially homogeneous, with the the velocity space distributions of the core and beam ions initialized with bi-Maxwellian and drifting bi-Maxwellian distributions with drift velocities, temperatures, densities, and anisotropies  given in Table~\ref{model_param:tab} and illustrated in Figure~\ref{vxvz_a_p_pv2av1.4Ap2Aa2betac0.214betab0.858}.



\newpage



\end{document}